\documentclass[traditabstract]{aa} 
\usepackage{txfonts}
\usepackage{graphicx}
\usepackage{natbib}

\newcommand{\jshort}{\mbox{\object{J1832}}}
\newcommand{\jmid}{\mbox{\object{J1832.4-1627}}}
\newcommand{\jlong}{\mbox{\object{J183221.56-162724.25}}}

\newcommand{\exhya}{\mbox{\object{EX\,Hya}}}
\newcommand{\vsgr}{\mbox{\object{V1223\,Sgr}}}
\newcommand{\vv}{\mbox{\object{V2400\,Oph}}}

\newcommand{\iphas}{\mbox{\object{IPHAS\,J0627}}}
\newcommand{\kic}{\mbox{\object{KIC\,5608384}}} 
\newcommand{\halp}{H$\alpha$}

\newcommand{\heii}{\ion{He}{ii}$\lambda$4686}

\newcommand{\phirot}{$\phi_\mathrm{65}$}
\newcommand{\phirotc}{$\phi_\mathrm{65,c}$}
\newcommand{\pobs}{$P_\mathrm{obs}$}
\newcommand{\po}{$P_\mathrm{orb}$}
\newcommand{\ps}{$P_\mathrm{spin}$}

\newcommand{\pspo}{$P_\mathrm{spin}/P_\mathrm{orb}$}

\newcommand{\ebmv}{$E_\mathrm{B-V}$}

\newcommand{\logl}{log($L_2/L_\odot$)}

\newcommand{\ri}{$r\!-\!i$}
\newcommand{\rip}{$(r\!-\!i)_\mathrm{P1}$}

\newcommand{\nh}{$N_\mathrm{H}$}

\newcommand{\atoms}{H-atoms\,cm$^{-2}$}

\newcommand{\rwd}{$R_\mathrm{wd}$}
\newcommand{\rco}{$r_\mathrm{co}$}
\newcommand{\rci}{$r_\mathrm{ci}$}

\newcommand{\racc}{$r_\mathrm{acc}$}
\newcommand{\rmag}{$r_\mathrm{mag}$}
\newcommand{\ralf}{$r_\mathrm{alf}$}
\newcommand{\tmag}{$t_\mathrm{mag}$}

\newcommand{\tkep}{$t_\mathrm{kep}$}
\newcommand{\tvis}{$t_\mathrm{vis}$}

\newcommand{\teffs}{$T_\mathrm{eff,2}$}

\newcommand{\ergs}{erg\,cm$^{-2}$s$^{-1}$}

\newcommand{\cms}{cm\,s$^{-1}$}
\newcommand{\ten}[2]{#1\,\times\!10^{#2}}

\newcommand{\rsun}{$R_\odot$}
\newcommand{\msun}{$M_\odot$}
\newcommand{\lsun}{$L_\odot$}
\newcommand{\msunyr}{$M_{\odot}$yr$^{-1}$}

\begin{document}
\title{\jmid, the first eclipsing stream-fed intermediate polar}

\author{
Beuermann, K. \inst{1} \and 
Breitenstein, P. \inst{2} \and 
Schwab, E. \inst{3}
}

\institute{ Institut f\"ur Astrophysik, Georg-August-Universit\"at,
  Friedrich-Hund-Platz 1, 37077 G\"ottingen, Germany,\\
  e-mail: beuermann@astro.physik.uni-goettingen.de
  \and Pascal Gymnasium, Uppenkampstiege 17, 48147 M\"unster,
  Astronomy and Internet in M\"unster (AiM),
  Dorotheenstr. 17, 48145 M\"unster, Germany, e-mail:
  p-breitenstein@online.de \and
  GSI Helmholtzzentrum f\"ur Schwerionenforschung GmbH, Planckstr.\,1,
  64291 Darmstadt, Germany,
  Volunteer for ESA/ESOC, Robert-Bosch Str. 5, 64293 Darmstadt, Germany,
  e-mail: e.schwab@gsi.de }
\date{Received 18 June 2021; accepted 21 Oct 2021}

\authorrunning{K. Beuermann et al.} 
\titlerunning {First eclipsing stream-fed intermediate polar.}
  
\abstract{We present a photometric study of the newly discovered
  eclipsing intermediate polar \jlong\ (in short \jshort) with an
  orbital period of 8.87\,hr. The system features a box-like deep
  eclipse with a full width at 50\% depth of $1970\pm2$\,s and a
  large-amplitude coherent pulsation with \pobs$\,=\!65.18$\,min,
  which represents either the synodic (beat) period or the spin period
  of the white dwarf (WD). The period ratio is either
  \pspo$\,=\,$0.1091 or 0.1225, respectively. The eclipsed light
  originates almost entirely from the two accretion spots and columns
  on the WD, with characteristics indicative of pole flipping. There
  is no evidence for an accretion disk, and we identify \jshort\ as
  the first deeply eclipsing stream-fed intermediate polar. Our
  $grizy$ photometry in eclipse yielded an $i$-band AB magnitude of
  the Roche-lobe-filling secondary star of 18.98(3), an extinction
  \ebmv$\,=\!0.54\!\pm\!0.17$, and a spectral type $\sim\,$K6. Dynamic
  models, fitting the photometry, limit the distance to between 1270
  and 2500\,pc for masses of the secondary star, $M_2$, between 0.16
  and 1.0\,\msun, well within the Gaia EDR3 confidence limits. Employing a
  luminosity selection inspired by binary population studies yields a
  mean $M_2\!=\!0.32$\,\msun\ with a 2$\sigma$ upper limit of
  0.60\,\msun\ and a mean distance $d\!=\!1596$\,pc with a 2$\sigma$
  upper limit of 1980\,pc. The secondary star is located in its
  Hertzsprung-Russell diagram at a mean $T_\mathrm{eff,2}\!=\!4120$\,K
  and log$(L_2/L_\mathrm{\odot})\!=\!-0.92$, from where the binary can
  evolve into either a polar or an ultracompact binary with a highly
  magnetic primary. The system displays a variable accretion rate,
  lapses repeatedly into short-lived low states of negligible
  accretion, and currently displays an orbital period that decreases
  on a timescale of $\tau\!\sim\!\ten{3}{5}$\,yr. X-ray observations,
  optical spectroscopy, and spectropolarimetry have a high potential
  for studies of the properties of \jshort\ as an individual object
  and of stream-fed accretion in general.}

\keywords{Stars: cataclysmic variables -- Stars: binaries: close --
  Stars: evolution -- Stars: individual: J1832.4-1627}

\maketitle


\section{Introduction}

Magnetic cataclysmic variables (CVs) contain an accreting white dwarf
(WD) with a surface field strength in the MG regime fed by a low-mass
secondary star. They come in two flavors, the synchronized polars (or
AM Herculis stars) and the non-synchronized intermediate polars (IPs).
Polars lack an accretion disk, and the WD accretes via a
quasi-stationary stream that guides the matter from the inner
Lagrangian point, $L_1$, more or less directly to the WD. In most IPs,
on the other hand, the WD accretes via the intermediary of an
accretion disk with an inner bound defined by the equality of viscous
and magnetic stresses at the radius of the WD's magnetosphere. The
difference reflects the typically larger orbital periods and system
dimensions of IPs and, perhaps, the typically lower magnetic moments
of their primary stars. An intriguing, yet controversial, idea
suggested an evolutionary link between the two, identifying IPs as the
progenitors of polars
\citep[e.g.,][]{kingetal85,hameuryetal86,wickramasingheetal91,kinglasota91,patterson94,nortonetal04,nortonetal08,southworthetal07}.
The proposition postulates the existence of non-synchronized
counterparts to polars, which synchronize and appear as polars in
their further evolution toward shorter orbital periods. An IP,
approaching such a scenario, might have a freely spinning WD with a
sufficiently strong magnetic moment to avoid the formation of a disk
and accrete instead from a stream that alternately feeds its two
poles, a situation referred to as pole flipping. Unfortunately, no
such IP has yet been found. The process of synchronization has been
studied, however, in asynchronously rotating polars that were already
synchronized and had temporarily lost synchronism by some event, such
as a nova outburst, and are currently re-synchronizing. They display
spin and orbital periods that differ by only $\sim\!1$\%, which
corresponds to the situation in an IP just preceding
synchronization. Well-studied cases include V1500\,Cyg\,=\,Nova Cyg
1975, which seems to have acquired synchronism again after 40 years
\citep{harrisoncampbell16}, and the still asynchronous systems BY\,Cam
\citep{honeycuttkafka05}, V1432\,Aql \citep{littlefieldetal15}, and
CD\,Ind \citep{littlefieldetal19}. Because of the near equality of
spin and orbital periods, the synodic (beat) periods are long, between
7 and about 62 days. As a consequence, the phenomenology of
asynchronous polars is generally rather complex, more complex, in
fact, than expected for a stream-fed IP that is still far from
synchronism.
  
Despite intense searches, only one secure candidate of a stream-fed IP
has so far been discovered, \vv, which, unfortunately, is not
particularly well suited for studying the conjectured IP-polar
connection. It is seen at an inclination of only $\sim\!10\degr$, and
consequently its system parameters are not well known
\citep{buckleyetal95,buckleyetal97,hellierbeardmore02,joshietal19}. Other
IPs display a mixture of disk-fed and stream-fed accretion, showing
pulsations at both the spin and the synodic period that are produced
by the phenomenon of stream overflow. Part of the stream that impacts
on the outer disk manages to flow over the disk and carry its angular
momentum directly into the WD's magnetosphere \citep{lubow89}. Still
others, such as TX\,Col, usually dominated by disk-fed accretion, show
stream-fed intervals after losing and before reestablishing the disk
\citep{littlefieldetal21}, which is of interest in itself as it allows
the process of disk formation to be observed.

Eclipsing systems offer obvious advantages for in-depth studies of the
system parameters. They permit reliable measurements of the component
masses, allow the usually faint secondary star to be detected in
eclipse, and provide insight into the structure of the accretion
region because practically the entire surface of the WD comes into
view as it rotates. Unfortunately, the number of eclipsing IPs is
exceedingly small. Of the 117 IPs and IP candidates in the last
version of the \citet{ritterkolb03} catalog (final online version
7.24, 2016), only four show a deep eclipse. These are, with one addition,
the classical systems DQ Her (Nova Her 1934) and XY Ari,
V902~Mon=IPHAS\,J0627 \citep{aungwer12}, V597~Pup, the now faint Nova
Puppis 2007 \citep{warnerwoudt09}, and Nova\,Sco~1437 \citep[][and
  references therein]{potterbuckley18}. Alas, all five contain
luminous accretion disks.

A feature shared by some IPs and many polars are irregular long-term
brightness variations, including states of vanishing accretion,
referred to as low states. In the usual high states, the accretion
luminosity outshines the stellar components, and, in many systems, the WD
becomes detectable only in a low state. By no means do all IPs provide this
opportunity. The well-studied 13\,mag disk-fed system EX Hya, for example,
exhibits rare outbursts but has not lost its bright disk since
observations started more than 60 yr ago (see the AAVSO long-term
light curve\footnote{https://www.aavso.org/vsots$\_$exhya}).  As a
consequence, the properties of both stellar components are still not
well known
\citep{eisenbartetal02,belleetal03,hoogerwerfetal04,beuermannreinsch08,beuermannreinsch13}.

In this paper we present a photometric study of \jlong\ (henceforth
\jshort), the first deeply eclipsing stream-fed IP that has it all:
(i) Its orbital period of \mbox{8.87\,hr} and a coherent
large-amplitude 65-minute pulsation qualify it as an IP; (ii) it
showed no evidence for an accretion disk in three consecutive
observing seasons, suggesting that this is a long-lasting feature;
(iii) the emitted light, apart from that of the secondary star,
originates from the immediate vicinity of the WD, probably the tall
accretion columns at both poles; (iv) the entire central region
suffers a deep eclipse by the secondary star; (v) ingress and egress
are fast and unresolved at our time resolution of about 30\,s; (vi)
the evidence for pole flipping supports its identification as a
stream-fed IP; and (vii) it lapsed repeatedly into low states,
offering the prospect for orbital phase-resolved studies of the
secondary star and the WD in spectral regions where either one
dominates. No object similar to short\ is currently known.

\begin{figure}[t]
\hspace{1.0mm}\includegraphics[width=88.0mm,angle=0,clip]{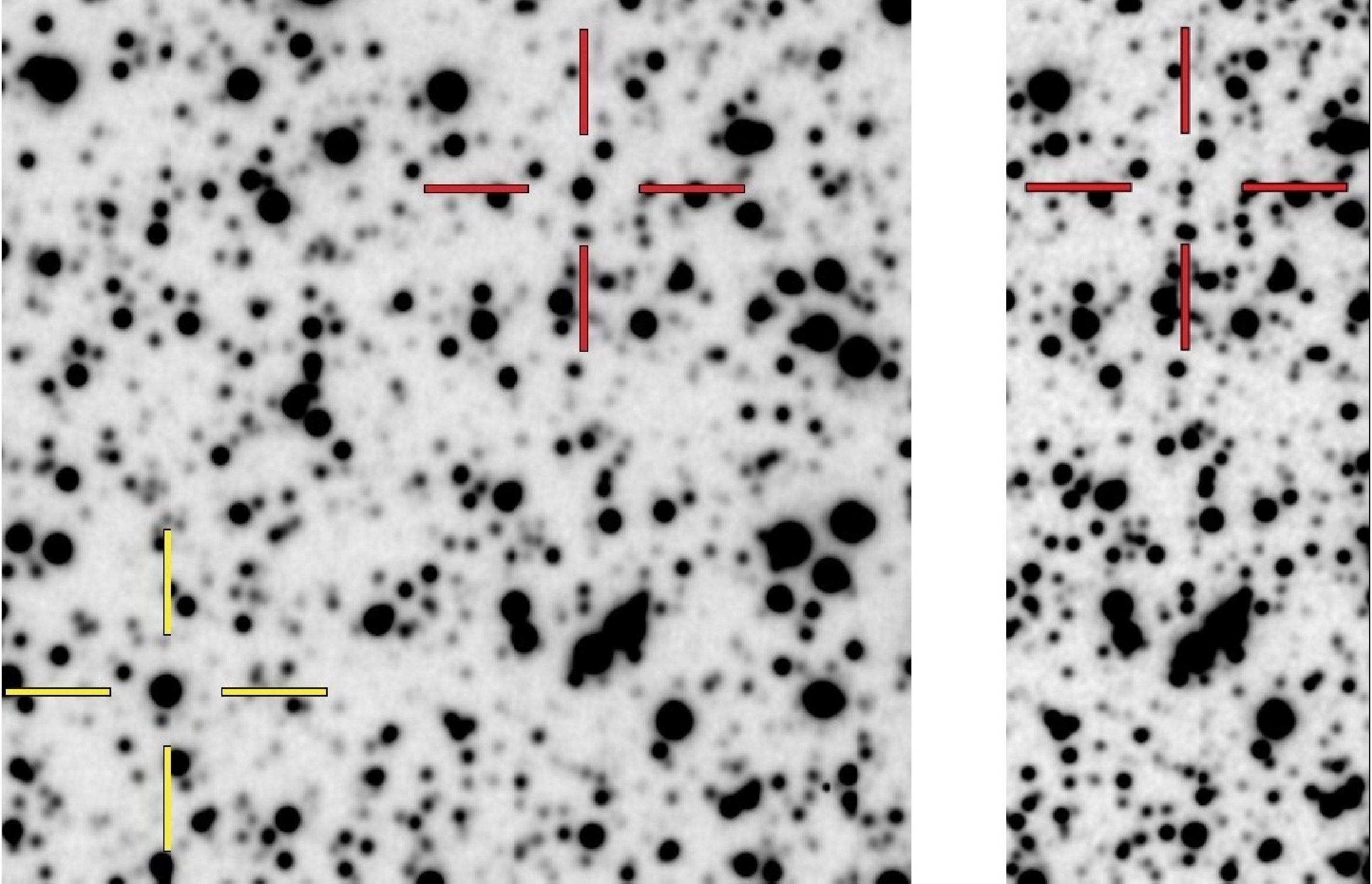}
\caption[chart]{Finding chart for \jlong. The $r$-band images were
  taken on 12 May 2021 with MuSCAT3 on the FTN. N is up, E to the
  left, and sizes are 80\arcsec$\times$80\arcsec\ and
  30\arcsec$\times$80\arcsec. The target and the comparison star are
  indicated by the red and yellow crosses, respectively. The right
  panel shows the target in the eclipse.}
\label{fig:fc}
\end{figure}

The new IP was discovered on 26 July 2019 by E. Schwab as a side
product of the search for the lost comet P/2012 K3 (Gibbs) using the
80\,cm Schmidt telescope of the German-Spanish Astronomical Center at
Calar Alto (CA), Spain, and described by
\citet{schwabbreitenstein19}. It received the AAVSO\footnote{American
  Association of Variable Star Observers} Unique Identifier
000-BNG-512. It was previously (and incorrectly) automatically
classified in the Pan-STARRS 3$\pi$ survey \citep{sesaretal17} as an
RR Lyr star. It carries the Gaia Data Release 2 (DR2) identifier
4102856333775127296 and is located at RA,Dec. (2000)
18\,32\,21.56,$-$16\,27\,24.2 ($l,b=16\fdg1351,-3\fdg3367$) in the
Milky Way (Fig.\,\ref{fig:fc}). The Gaia Early Data Release 3 (EDR3)
lower and upper confidence limits for the geometric distance are 1254
and 2847\,pc, respectively
\citep{bailerjonesetal21}\footnote{http://vizier.u-strasbg.fr/viz-bin/VizieR?-source=I/352}.
We present a photometric study of the object, including an analysis of
phase-resolved light curves taken in white light (WL) and multicolor
$grizy$ light curves taken through the eclipse. The extinction as
measured by us is \ebmv$\,=\!0.54\!\pm\!0.17$. Measurements of the
brightness and spectral type of the secondary star allowed us to
restrict the system parameters.

\section{Observations}
\label{sec:obs}

A total of 3551 WL images were taken on 32 nights in 2019, 2020, and
2021 with the 80\,cm Schmidt telescope at the CA Observatory, Spain,
equipped in 2019 with an SBIG ST\,10-XME charge-coupled device camera
and in 2020 and 2021 with a Proline PL23042 camera. The 2019 data were
described by \citet{schwabbreitenstein19}. Here, we summarize the
combined results of the 2019, 2020, and 2021 campaigns and report
updated ephemerides for the orbital motion and the rotation of the WD.
All CA data were taken with a clear filter in WL and exposed for
60\,s. Additional long runs in WL were performed with the MONET/South
telescope at the Sutherland site of the South African Astronomical
Observatory (SAAO).
Shorter observations that covered preferentially the eclipse were
performed with telescopes of the Las Cumbres Observatory (LCO), namely
the 2-meter Faulkes Telescope South (FTS) and the 1-meter telescope at Siding
Spring, Australia, the 2-meter Faulkes Telescope North (FTN) at the
Haleakala Observatory, Hawaii, and the 1-meter telescope at the McDonald
Observatory, Texas, USA. These runs were taken in the $grizy$ filters
on the Pan-STARRS1 (P1) system \citep{tonryetal12} or in WL with
exposure times mostly of 40\,s.  Table~\ref{tab:log} provides a log of
the observations.

\begin{table}[t]
\begin{flushleft}
\caption{Journal of time-resolved photometric observations of \jshort.}
\begin{tabular}{@{\hspace{1.0mm}}l@{\hspace{1.5mm}}r@{\hspace{0.0mm}}c@{\hspace{0.0mm}}r@{\hspace{1.5mm}}c@{\hspace{1.5mm}}c@{\hspace{1.5mm}}c@{\hspace{1.5mm}}c}\\[-4ex]
  \hline\hline                                                                     \\[-1.5ex]
  Telescope        & Year & No.    & No.    & Band    & Exp.       & Plate  & Tel.  \\
                   &      & nights & exp    &         & (s)        & scale  &       \\[0.5ex]
  \hline                                                              &            \\[-1ex]
  CA Schmidt       & 2019 & 12     & 1410   &  WL     & 60         &   0.74 & (1)   \\       
  SAAO MONET/S     & 2019 & ~~3    &  270   &  WL     & 90,180     &   0.37 & (2)   \\
  LCO FTS          & 2019 & 11     &  138   &  WL,r   & 60,90,120  &   0.30 & (3)   \\
  LCO 1\,m McD     & 2019 & ~~2    &   19   &  WL,r   & 60         &   0.78 & (5)   \\
  CA Schmidt       & 2020 & 14     & 1410   &  WL     & 60         &   1.29 & (1)   \\
  LCO 1\,m SS      & 2020 & ~~1    &   16   &  r      & 60         &   0.78 & (6)   \\
  LCO FTN,FTS      & 2020 & 14     &  228   &  grizy  & 40,60,120  &   0.30 & (4,3) \\
  CA Schmidt       & 2021 & ~~6    &  731   &  WL     & 60         &   1.29 & (1)   \\
  LCO FTN,FTS      & 2021 & ~~6    &  183   & griz,WL & 40,50,60   &   0.30 & (4,3) \\
  LCO 1\,m McD,SS  & 2021 & ~~2    &  142   &  WL     & 20,50      &   0.78 & (5,6) \\
  LCO 1\,m CT,SAAO & 2021 & ~~5    &  155   &  g      & 60         &   0.78 & (7,8) \\
  [1.0ex]
 \hline\\
\end{tabular}\\[-1.0ex]
\footnotesize{ WL = white light. Plate scale is in \arcsec/pix. See
  also footnote to Table~3.}
\label{tab:log}
\end{flushleft}
\vspace{-4mm}
\end{table}

\begin{table}[t]
\begin{flushleft}
\caption{Mid-eclipse times of \jshort.}
\begin{tabular}{@{\hspace{0.0mm}}r@{\hspace{3.0mm}}r@{\hspace{3.0mm}}r@{\hspace{3.0mm}}r@{\hspace{3.0mm}}r@{\hspace{3.0mm}}c}\\[-2ex]
\hline\hline                                                  \\[-1.5ex]
Cycle& Observed    & Calculated  & Error   & O-C      & Tel.  \\
      &TDB 240000+ & TDB 240000+ & (days)  & (days)   &       \\[0.5ex]
  \hline                                                      \\[-1ex]                                                   
   0 & 58691.53623 & 58691.53666 & 0.00006 & -0.00043 & (1)   \\            
  16 & 58697.44818 & 58697.44847 & 0.00007 & -0.00029 & (1)   \\            
  89 & 58724.42085 & 58724.42113 & 0.00006 & -0.00028 & (1)   \\            
 105 & 58730.33282 & 58730.33295 & 0.00011 & -0.00013 & (2)   \\            
 140 & 58743.26482 & 58743.26504 & 0.00009 & -0.00022 & (2)   \\            
 197 & 58764.32580 & 58764.32588 & 0.00006 & -0.00008 & (1)   \\            
 812 & 58991.56180 & 58991.56127 & 0.00006 &  0.00052 & (1)   \\            
 831 & 58998.58187 & 58998.58155 & 0.00005 &  0.00032 & (1)   \\            
 881 & 59017.05611 & 59017.05597 & 0.00008 &  0.00013 & (3)   \\            
 912 & 59028.51013 & 59028.51011 & 0.00006 &  0.00002 & (1)   \\            
 913 & 59028.87975 & 59028.87960 & 0.00006 &  0.00015 & (4)   \\            
 919 & 59031.09682 & 59031.09653 & 0.00014 &  0.00029 & (5)   \\            
 962 & 59046.98473 & 59046.98454 & 0.00009 &  0.00019 & (4)   \\            
 970 & 59049.94079 & 59049.94044 & 0.00007 &  0.00035 & (4)   \\            
1047 & 59078.39118 & 59078.39105 & 0.00011 &  0.00012 & (1)   \\            
1050 & 59079.49998 & 59079.49952 & 0.00006 &  0.00046 & (1)   \\            
1105 & 59099.82179 & 59099.82138 & 0.00018 &  0.00041 & (4)   \\            
1181 & 59127.90283 & 59127.90250 & 0.00021 &  0.00033 & (5)   \\            
1704 & 59321.14464 & 59321.14496 & 0.00007 & -0.00032 & (4)   \\            
1726 & 59329.27385 & 59329.27370 & 0.00018 &  0.00015 & (5)   \\            
1746 & 59336.66327 & 59336.66347 & 0.00006 & -0.00020 & (1)   \\            
1763 & 59342.94466 & 59342.94477 & 0.00009 & -0.00012 & (3)   \\            
1766 & 59344.05304 & 59344.05324 & 0.00005 & -0.00020 & (4)   \\            
1773 & 59346.63959 & 59346.63966 & 0.00006 & -0.00007 & (1)   \\            
1774 & 59347.00911 & 59347.00915 & 0.00005 & -0.00003 & (4)   \\            
1800 & 59356.61574 & 59356.61585 & 0.00007 & -0.00011 & (1)   \\    
1842 & 59372.13429 & 59372.13436 & 0.00005 & -0.00007 & (6)   \\[1.0ex]     
\hline\\
\end{tabular}\\[-1.0ex]
\footnotesize{See footnote to Table~3 for information on the telescopes.}
\label{tab:orb}
\end{flushleft}
\vspace{-5mm}
\end{table}

\begin{table}[t]
\begin{flushleft}
\caption{Minima of the 65-minute pulsation of \jshort.}
\vspace{-1.5mm}
\begin{tabular}{@{\hspace{1.0mm}}r@{\hspace{3.0mm}}r@{\hspace{3.0mm}}r@{\hspace{3.0mm}}r@{\hspace{3.0mm}}r@{\hspace{3.0mm}}c}\\[-1ex]
  \hline\hline                                                    \\[-1.5ex]
Cycle  & Observed    & Calculated  & Error     & O-C     & Tel. \\
       & TDB 240000+ & TDB 240000+ & (days)    & (days)  &      \\[0.5ex]
  \hline                                                          \\[-1ex]                                                   
    0  & 58691.44610 & 58691.44620 & 0.00180 & -0.00010 & (1) \\       
   22  & 58692.44140 & 58692.44200 & 0.00120 & -0.00060 & (1) \\       
   23  & 58692.48820 & 58692.48726 & 0.00230 &  0.00094 & (1) \\       
   65  & 58694.38780 & 58694.38833 & 0.00120 & -0.00053 & (1) \\       
   66  & 58694.43610 & 58694.43360 & 0.00180 &  0.00250 & (1) \\       
   87  & 58695.38200 & 58695.38413 & 0.00340 & -0.00213 & (1) \\       
   88  & 58695.42850 & 58695.42939 & 0.00120 & -0.00089 & (1) \\       
   89  & 58695.47490 & 58695.47466 & 0.00230 &  0.00024 & (1) \\       
  110  & 58696.42290 & 58696.42519 & 0.00230 & -0.00229 & (1) \\       
  111  & 58696.46830 & 58696.47046 & 0.00230 & -0.00216 & (1) \\       
  132  & 58697.42280 & 58697.42099 & 0.00230 &  0.00181 & (1) \\       
  133  & 58697.46910 & 58697.46626 & 0.00230 &  0.00284 & (1) \\       
  529  & 58715.38840 & 58715.39064 & 0.00340 & -0.00224 & (1) \\     
  530  & 58715.43670 & 58715.43591 & 0.00180 &  0.00079 & (1) \\       
  531  & 58715.48490 & 58715.48117 & 0.00230 &  0.00373 & (1) \\       
  727  & 58724.35270 & 58724.35284 & 0.00180 & -0.00014 & (1) \\       
  728  & 58724.39620 & 58724.39810 & 0.00230 & -0.00190 & (1) \\       
  857  & 58730.23700 & 58730.23711 & 0.00250 & -0.00011 & (2) \\       
  858  & 58730.28220 & 58730.28237 & 0.00250 &  0.00017 & (2) \\       
 1123  & 58742.27660 & 58742.27723 & 0.00190 & -0.00063 & (2) \\
 1124  & 58742.32310 & 58742.32249 & 0.00190 &  0.00061 & (2) \\
 1125  & 58742.36950 & 58742.36775 & 0.00130 &  0.00175 & (2) \\
 1146  & 58743.31480 & 58743.31829 & 0.00360 & -0.00349 & (2) \\
 1611  & 58764.36580 & 58764.36587 & 0.00230 & -0.00007 & (1) \\
 1632  & 58765.31570 & 58765.31640 & 0.00180 & -0.00070 & (1) \\
 1633  & 58765.36110 & 58765.36167 & 0.00230 & -0.00056 & (1) \\
 6631  & 58991.58810 & 58991.58916 & 0.00180 & -0.00106 & (1) \\
 6697  & 58994.57600 & 58994.57656 & 0.00140 & -0.00056 & (1) \\
 6698  & 58994.62120 & 58994.62182 & 0.00180 & -0.00062 & (1) \\
 6741  & 58996.56690 & 58996.56816 & 0.00140 & -0.00126 & (1) \\
 6785  & 58998.56010 & 58998.55976 & 0.00140 &  0.00034 & (1) \\
 7292  & 59021.51070 & 59021.50840 & 0.00280 &  0.00230 & (1) \\
 7293  & 59021.55590 & 59021.55367 & 0.00280 &  0.00223 & (1) \\
 7314  & 59022.50360 & 59022.50420 & 0.00140 & -0.00060 & (1) \\
 7315  & 59022.54960 & 59022.54947 & 0.00140 &  0.00013 & (1) \\
 7336  & 59023.49970 & 59023.50000 & 0.00180 & -0.00030 & (1) \\
 7337  & 59023.54270 & 59023.54527 & 0.00180 & -0.00257 & (1) \\
 7402  & 59026.48670 & 59026.48740 & 0.00180 & -0.00070 & (1) \\
 7403  & 59026.53190 & 59026.53267 & 0.00180 & -0.00077 & (1) \\
 7404  & 59026.57790 & 59026.57793 & 0.00300 &  0.00003 & (1) \\
 7446  & 59028.48290 & 59028.47900 & 0.00300 &  0.00390 & (1) \\
 7447  & 59028.53040 & 59028.52426 & 0.00300 &  0.00614 & (1) \\
 7448  & 59028.57410 & 59028.56953 & 0.00180 &  0.00457 & (1) \\
 8550  & 59078.44940 & 59078.45002 & 0.00180 & -0.00062 & (1) \\
 8594  & 59080.44080 & 59080.44161 & 0.00180 & -0.00081 & (1) \\       
15070  & 59373.56875 & 59373.56872 & 0.00180 &  0.00003 & (1) \\   
16460  & 59436.48437 & 59436.48514 & 0.00180 & -0.00076 & (1) \\[1.0ex]     
 \hline\\                            
\end{tabular}\\[-1.0ex]
  \footnotesize{ (1) CA, 80 cm Schmidt, Calar Alto/Spain, (2) SAAO,
    MONET\,/\,S, 1.2~m, Sutherland/SA, (3) LCO, 2 m FTS, Siding
    Spring, (4) LCO, 2 m FTN, Hawaii, (5) LCO, 1 m McDonald/Texas, (6)
    LCO, 1 m Siding Spring}, (7) LCO, 1 m Cerro Tololo/Chile, (8) LCO,
  1 m SAAO.
  \label{tab:spin}
\end{flushleft}
\end{table}

\begin{figure}[t]
\includegraphics[height=89.0mm,angle=270,clip]{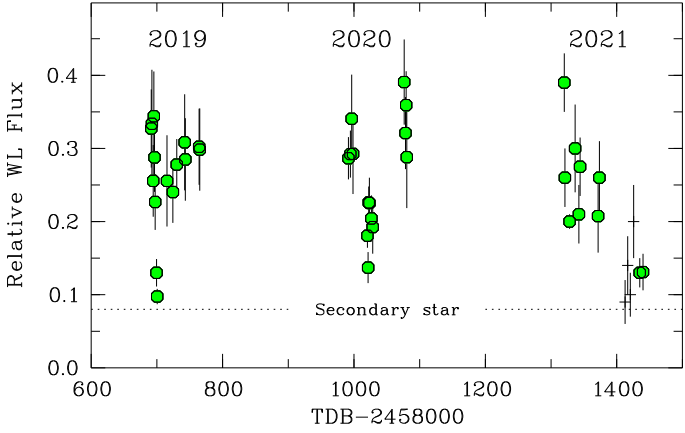}
\caption[chart]{Long-term light curve, showing nightly averages of the
  WL fluxes relative to comparison star C3 (green dots). The $g$-band
  fluxes of July 2021 (crosses) are adjusted to approximately match
  the WL fluxes. }
\label{fig:long}
\end{figure}

\begin{figure*}[t]
\includegraphics[height=89.0mm,angle=270,clip]{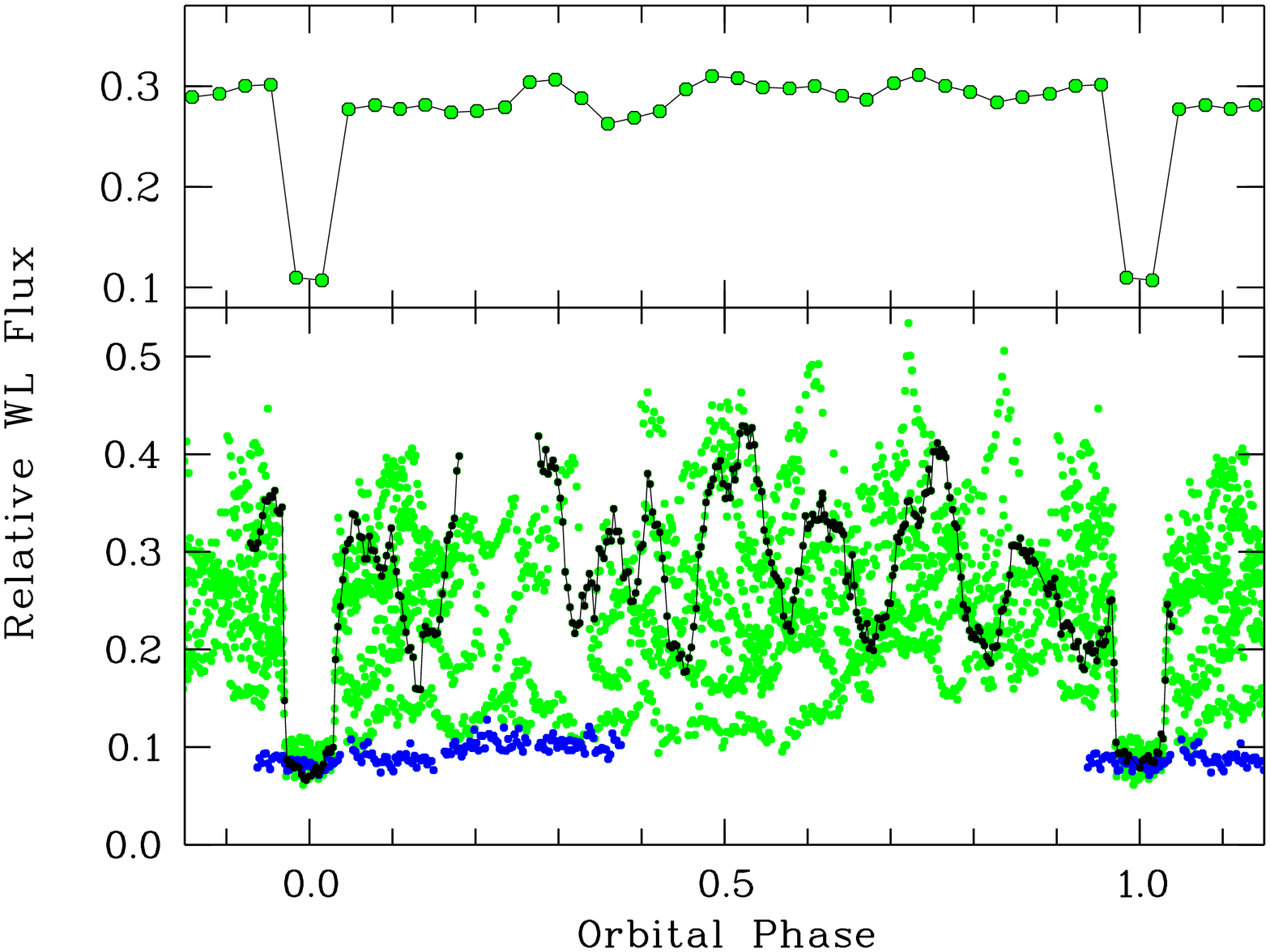}
\hfill
\includegraphics[height=89mm,angle=270,clip]{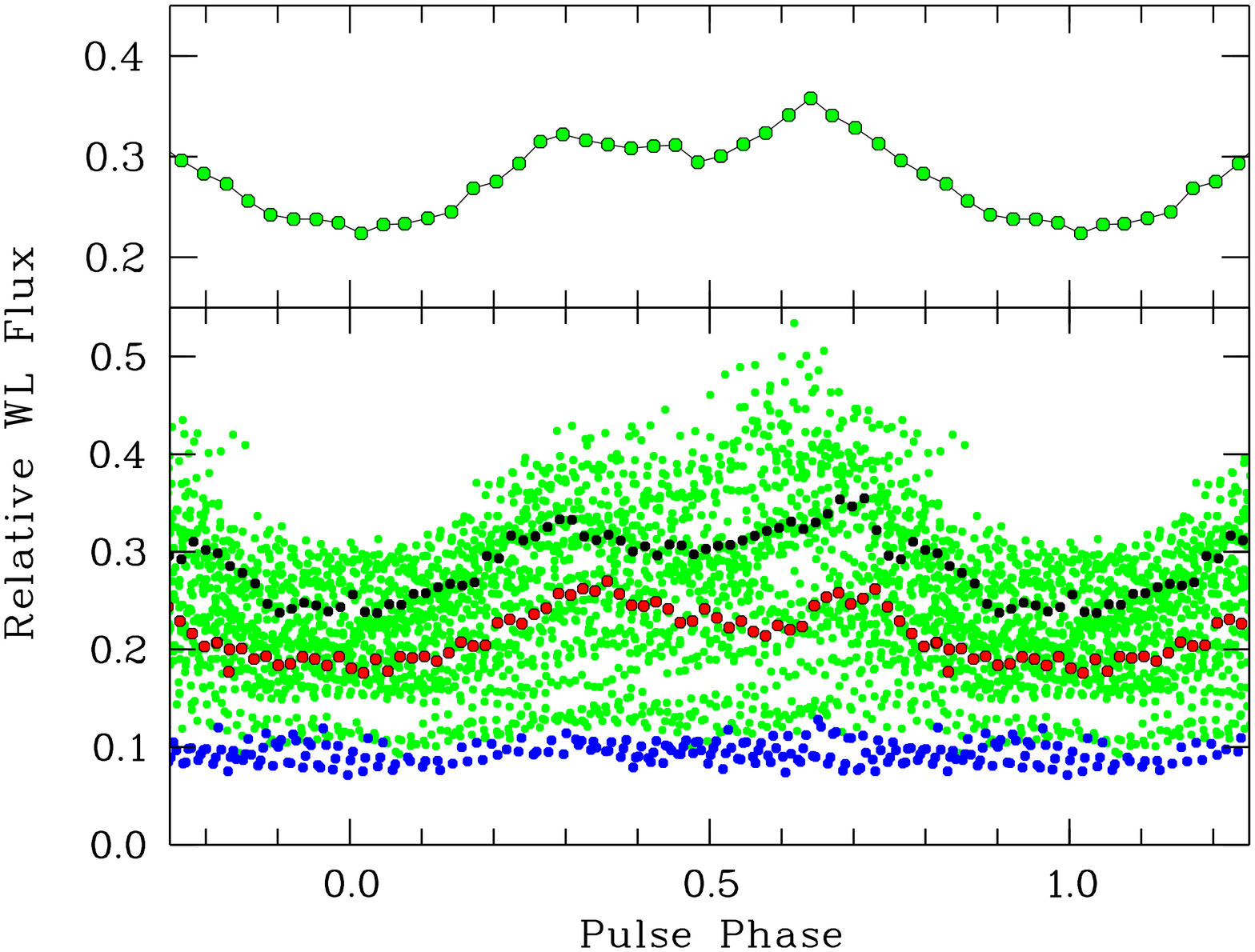}
\caption[chart]{\emph{Bottom panels: } Light curves in WL folded on
  the orbital period of Eq.~\ref{eq:orb1} (left) and on the pulse
  period of Eq.~\ref{eq:spin} (right). All individual relative WL
  fluxes are shown as green dots, and selected individual light curves are
  emphasized in black, red, and blue (see text). \emph{Top
      panels: }Mean light curves collected into 32 orbital bins
  (left) and pulse bins (right).}
\label{fig:lc}
\end{figure*}

\begin{figure*}[t]
\includegraphics[height=89mm,angle=270,clip]{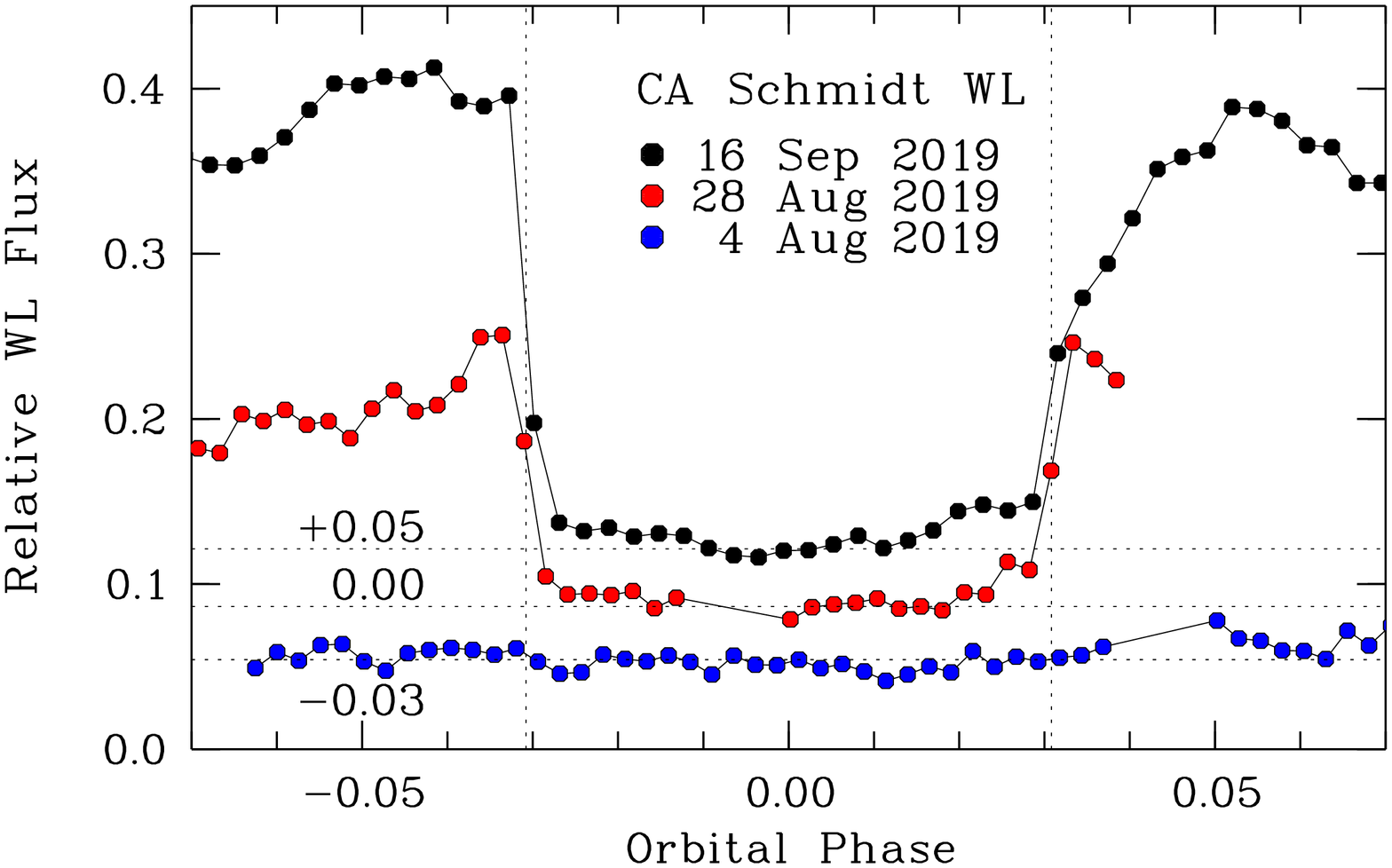}
\hfill
\includegraphics[height=89mm,angle=270,clip]{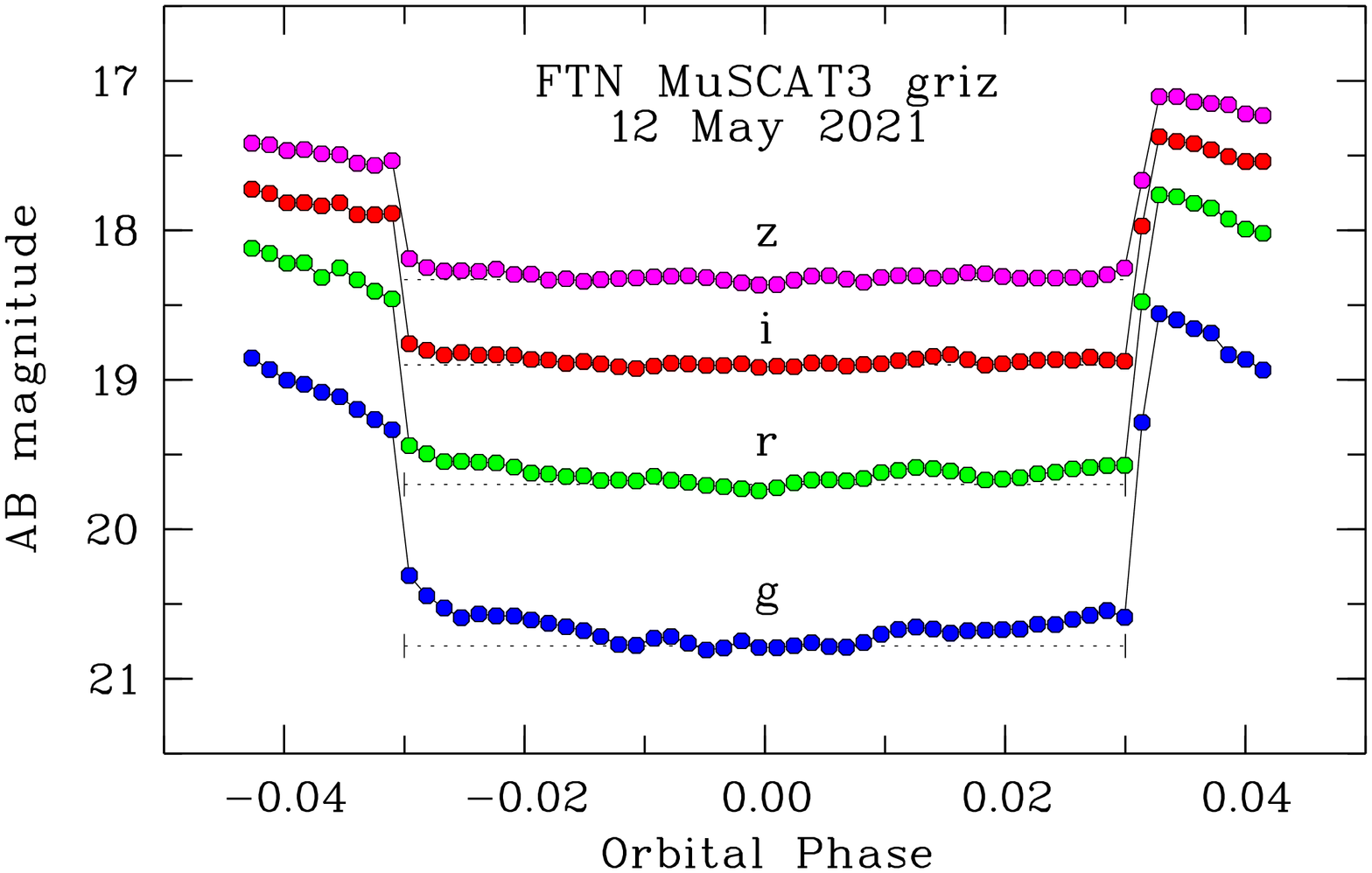}
\caption[chart]{\emph{Left: }Eclipse light curves in WL, taken
  with the CA Schmidt telescope, shifted vertically by the
  indicated amounts. \emph{Right: } Simultaneous eclipse light
  curves in the $griz$ filters, taken with the MuSCAT3 photometer on
  the FTN.  The dotted lines indicate the length and
  flux level of the totality.}
\label{fig:ecl}
\end{figure*}

All images were corrected for dark current and flat-fielded in the
usual way. We performed photometry relative to the comparison
star C3, which is located at RA,Dec. (2000) 18 32 24.118, -16 28 08.13,
or 57.2\arcsec\ SE of the target (Fig.~\ref{fig:fc}). It has
AB magnitudes on the P1 system $g\!=\!16.933(3), r\!=\!16.357(10),
i\!=\!16.107(8), z\!=\!15.931(7)$, and $y\!=\!15.897(7)$.
All times were measured in UTC and converted to
Barycentric Dynamical Time (TDB), using the tool provided by
\citet{eastmanetal10}\,\footnote{http://astroutils.astronomy.ohio-state.edu/time/},
which accounts also for the leap seconds.
As described in \citet{beuermannetal17}, the WL measurements
at the MONET/S telescope were tied into the standard $griz$ system. For most
incident spectra, the WL AB magnitude $w$ is within 0.1 mag of the
$r$-band magnitude. We assumed this calibration to be approximately
valid also for the CA WL measurements, which is supported by the
similar calibration of the P1 $w$-band \citep{tonryetal12}.

The experimental setups of the CA observations were tailored to comet
and asteroid searches and not optimized for studies of faint CVs in a
dense star field. In particular, the 2020 and 2021 setups with their
comparatively large plate scale (Table~\ref{tab:log}) led to spillover
of flux from the near neighbors visible in Fig.~\ref{fig:fc}. The
cameras at the MONET/S and the two Faulkes  telescopes had more
favorable plate scales and for these runs, a spillover problem from
the near neighbors at $\sim\!5$\arcsec\ did not exist. They were still
disturbed, however, by at least three faint companions at
$\sim\!1$\arcsec\ separation to the N, E, and S of the target, which
required a special effort in the photometric analysis.
The southern one is listed in the Gaia EDR3
\citep{gaia21} with $G\!=\!20.82$ at a separation of 1\farcs29. In the
Gaia high-state measurement, our target had $G\!=\!18.55$ and the
influence of the near neighbors was negligible, but in the eclipse
they may add noticeably to the measured flux, particularly in the
$g$-band, where the secondary star becomes faint.
A special effort is, therefore, required for reliable measurements of
the eclipse fluxes. Our method makes use of the variability of
the target: the difference image of equal exposures outside and
inside eclipse lacks all constant sources and contains a clean image
of the target in the selected filter. Its relative flux
$f_\mathrm{dif}$ was measured with respect to comparison star C3,
which has the physical flux $F_\mathrm{C3}$. We subtracted a fraction
$f_\mathrm{ecl}$ of the difference image from the eclipse image, such
that the result was as best as possible devoid of the
target. The target flux in eclipse is then
$f_\mathrm{ecl}\,f_\mathrm{dif}\,F_\mathrm{C3}$ and its error is
estimated from the uncertainty in establishing $f_\mathrm{ecl}$. The
method yielded background images in $g, r, i, z,$ and $y$ that
contained the near neighbors and illustrated the complexity of the
background at this position in the Milky Way.

\section{Temporal analysis}
\subsection{Light curves}
\label{sec:lc}

The largest body of light curves was taken with the CA Schmidt
telescope in WL (Table\,\ref{tab:log}). The source displayed
substantial long-term variability as shown in Fig.\,\ref{fig:long},
where the average relative WL fluxes are displayed versus time,
supplemented by $g$-band fluxes of July 2021. The ``error'' bars
represent the standard deviations of all individual flux measurements
of a given run. The peak-to-peak variations are another factor of
about two larger. Most of the time, the source was found at an average
WL flux of around 0.30 relative to the comparison star, which we take
to represent the high state of the source. In all three observing
seasons, the system lapsed into short-lived low or intermediate states
of accretion. It was found in a low state on $3-5$ August 2019 (JD
2458700), in an intermediate state between 20 and 27 June 2020 (JD
2459020-9028), and it oscillated between low and intermediate states
between 17 July and at least 9 August 2021. Since the source is
accreting, we associate these fluctuations with variations in the
accretion rate, though not necessarily the mass loss rate from the
secondary star.

The two panels of Fig.\,\ref{fig:lc} show the WL measurements of 2019
and 2020 phased on the 8.87\,hr orbital period and on the 65-minute
pulse period of Eqs.~\ref{eq:orb1} and \ref{eq:spin},
respectively. All individual flux measurements are displayed as green
dots and individual light curves are emphasized by other colors. In
the upper parts of both panels, we show the average light curves
collected into 32 phase bins. All light curves are characterized by
the deep eclipses of the WD by the secondary star and large-amplitude
pulsations that we relate to the WD rotation. Examples of the eclipse
light curves are shown in Fig.~\ref{fig:ecl}. The left panel presents
WL light curves at three different accretion levels (note that the
zero levels of the ordinate are displaced with respect to each other)
and the right one simultaneous high-state $griz$ light curves. The
remnant flux at the center of the eclipse is independent of the
accretion level and reaches down to that of the secondary star at the
center of the eclipse (Sects.~\ref{sec:sec1} and \ref{sec:incl}). We
can, therefore, exclude a grazing or partial eclipse as in EX Hya
\citep{beuermannosborne88,rosenetal88,allanetal98}. The WD itself is
faint compared with the accretion spots and columns, as evidenced by
the light curve of 4 August 2019, when the system was caught in a
state, when accretion had dropped to a trickle and the eclipse had
practically vanished (Figs.\,\ref{fig:lc} and \ref{fig:ecl}, blue
dots, and Sect.~\ref{sec:wd}). Examples of the large-amplitude
pulsations are shown in Fig.\,\ref{fig:lc}. The light curve in the
left panel (black dots) was pieced together from four individual runs
to illustrate the magnitude of the effect.

The rotational light curves in Fig.~\ref{fig:lc} show that the pulse
maxima are more or less double-peaked and the mean rotational light
curve in the upper right-hand panel suggests that this is a permanent
feature of \jshort. Transient double peaks were observed in the
photometric light curves of \vv\ and V1062\,Tau as well
\citep{buckleyetal95,lipkinetal04}. Permanent double peaks are
naturally produced in pole-flipping stream-fed accretors, as
demonstrated by the model calculations of \citet[][their
  Fig.\,3]{ferrariowickramasinghe99}.

\subsection{Is there an accretion disk?}
\label{sec:disk}

The mean orbital light curve in the left panel of Fig.\,\ref{fig:lc}
is essentially flat, with the remnant modulation related to individual
65-minute pulsations. In particular, it lacks the orbital hump expected
if there were a viscous disk with a locally puffed-up bulge.
Furthermore, the eclipse profile in Fig.~\ref{fig:ecl} lacks disk
characteristics as the wide rounded-off component observed in
\iphas\ \citep{aungwer12} or the V-shaped eclipse of
\kic\ \citep{yuetal19}. Any disk component is, furthermore, limited to
stay below the minima of the spin modulation. At our moderate temporal
resolution, the transitions into and out of eclipse stay largely
unresolved. For the large body of CA WL observations with
exposure times of 60 s, the formal average of the fitted transition
times between the contact points is $t_\mathrm{ine}\!=\!80\pm56$\,s.
For the best-resolved of our observations with 20\,s exposure and 9\,s
readout, the transition still occurs within one or two time bin.
This corresponds to a lateral extent in the orbital plane of
$2\pi\,a\,t_\mathrm{ine}/P_\mathrm{orb}\!\la\!10^9$\,cm.  While this is
entirely consistent with the extent of the accretion columns discussed
in Sect.~\ref{sec:orb}, even a minimal disk is more than ten times
larger. Its diameter would exceed two times the corotation radius
where $q\!=\!M_2/M_1\!\sim\!0.5$ was used for the mass ratio, \po\ is
the orbital period, and we used the observed pulse period in
lieu of the spin period \ps. Hence, a luminous disk cannot hide in
the rapid transitions in \jshort.

We now consider the sagging flux profile in the eclipses displayed in
Fig.~\ref{fig:ecl}. The decreasing flux in the first third and the
increasing flux in the last third of the eclipse indicate that an
extended source exists that suffers a delayed or incomplete
eclipse. In the WL light curve of 16 September 2019, 92\% of the
eclipsed flux at ingress disappear within a minute and the remaining
8\% (until eclipse center) within the next 15 min. In the $r$-band
light curve of 12 May 2021, 3\% of the eclipsed flux reappear between
eclipse center and the end of totality and the remaining 97\% within
the final minute. Hence, in WL or $r$, a few percent of the total
high-state flux are contributed by the extended source. The MuSCAT3
simultaneous $griz$-observations in the right panel of
Fig.~\ref{fig:ecl} suggest that this component is rather blue. Our
nonsimultaneous observations show that it is highly variable, in
particular in the $g$-band. It seems plausible to assign this emission
to the outer part of the magnetically guided and likely optically thin
accretion stream or to the ballistic stream between $L_1$ and its
impact on the magnetosphere, which are both prone to a partial
eclipse.  For the quoted corotation radius, the stream rises to
$\sim\!\ten{1.5}{10}$\,cm above the orbital plane, sufficient for a
time-dependent fraction of the stream to escape eclipse (see also
Sect.~\ref{sec:incl}). Such a component was seen in other magnetic CVs
as well: the long-period IP V902~Mon showed \halp\ emission that did
not completely disappear in eclipse \citep{worpeletal18}, and so did
the long-period CV \kic\ \citep{yuetal19}, and the polar HY~Eri
\citep{beuermannetal17}. Hence, emission from the accretion stream far
away from the WD is a natural candidate for the variable flux of
\jshort\ in eclipse. Alternatively, we cannot exclude that a minimal
faint disk or a ring of matter, circulating the WD near
\rco\ \citep{king93,kingwynn99,nortonetal08}, contributes a few
percent to the optical flux.
  
A final argument against the presence of a viscous accretion disk
follows from the low-state observation of $3-5$ August 2019.  On 4
August (blue light curve in the left panel of Fig.~\ref{fig:ecl}) a
viscous disk was absent. Had it been present on 1 August 2019, when
the source was still bright, it would have had to have disappeared
within 2 days, which is hard to reconcile with the viscous timescale
$t_\mathrm{visc}\!\sim\!R^2/(\alpha Hc_\mathrm{s})\!\ga\!20$\,d for
$R\!>\,$\rco, a relative disk height $H/R\!\la\!0.02$, a sound speed
of 10\,km/s, and a Shakura-Sunyaev parameter $\alpha\!<\!1$.  In
summary, we conclude that \jshort\ lacks a viscous accretion disk.

\subsection{Orbital period and ephemeris}
\label{sec:orb}

\begin{figure}[t]
\includegraphics[height=89mm,angle=270,clip]{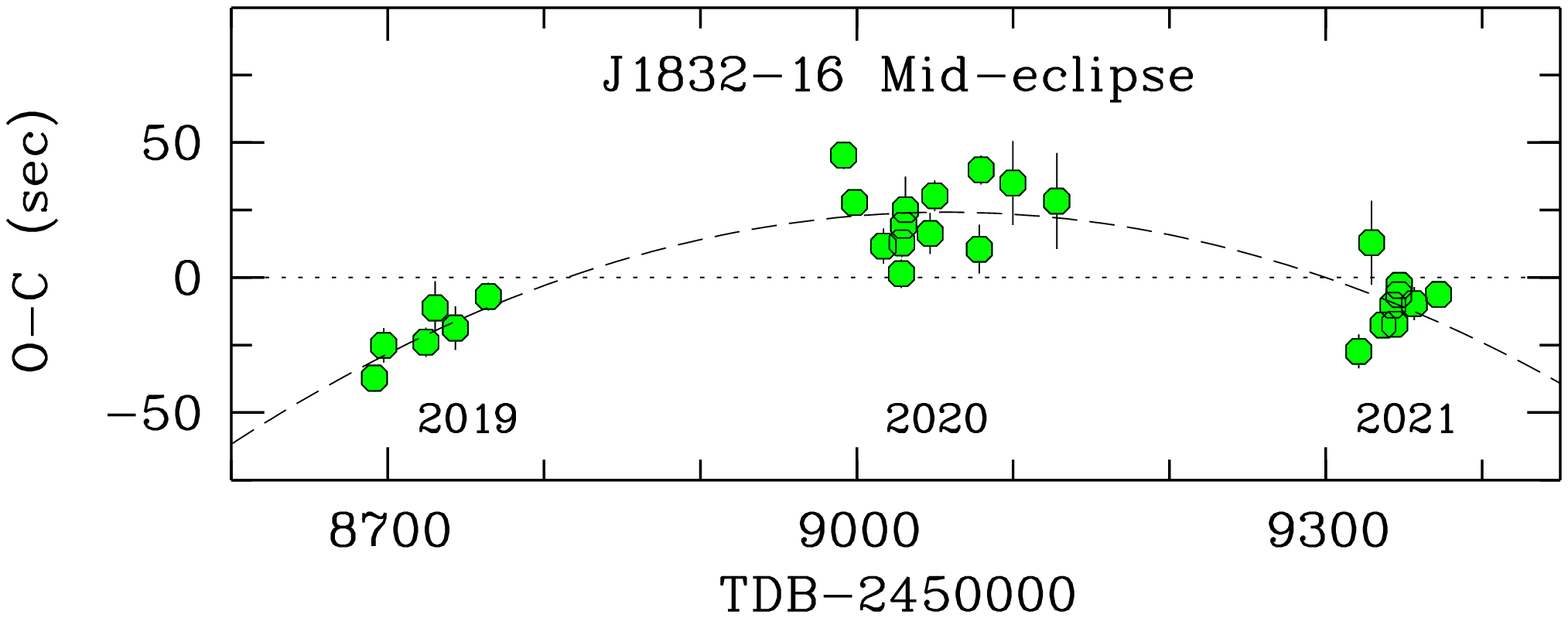}
\vspace{1.0mm}
\includegraphics[height=90.5mm,angle=270,clip]{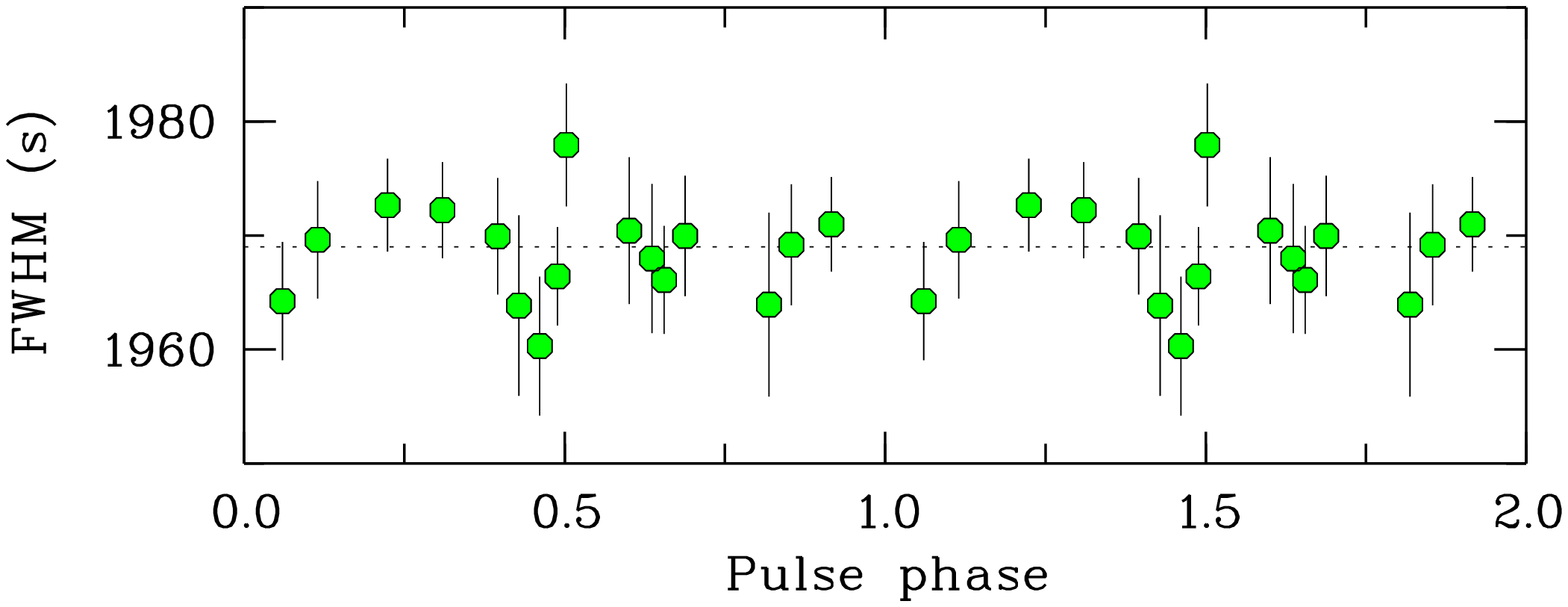}
\vspace{1.0mm}
\includegraphics[height=90mm,angle=270,clip]{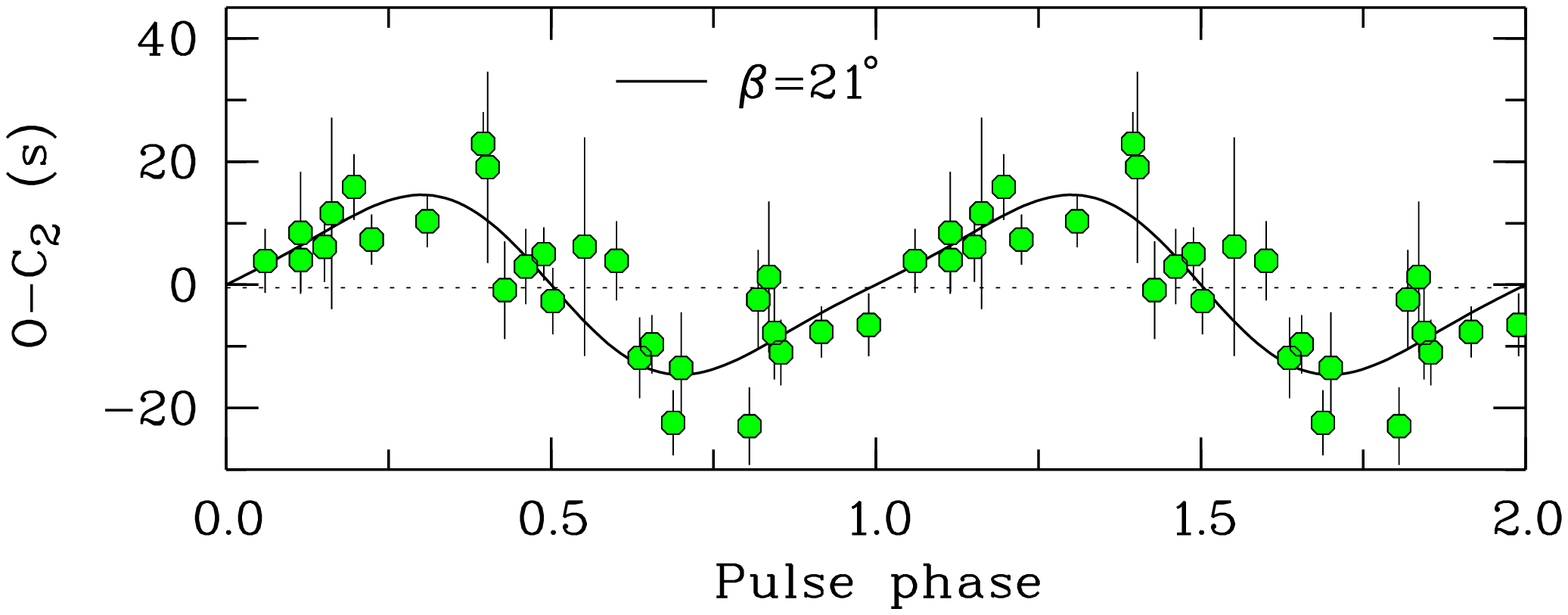}
\vspace{1.0mm}
\includegraphics[height=88mm,angle=270,clip]{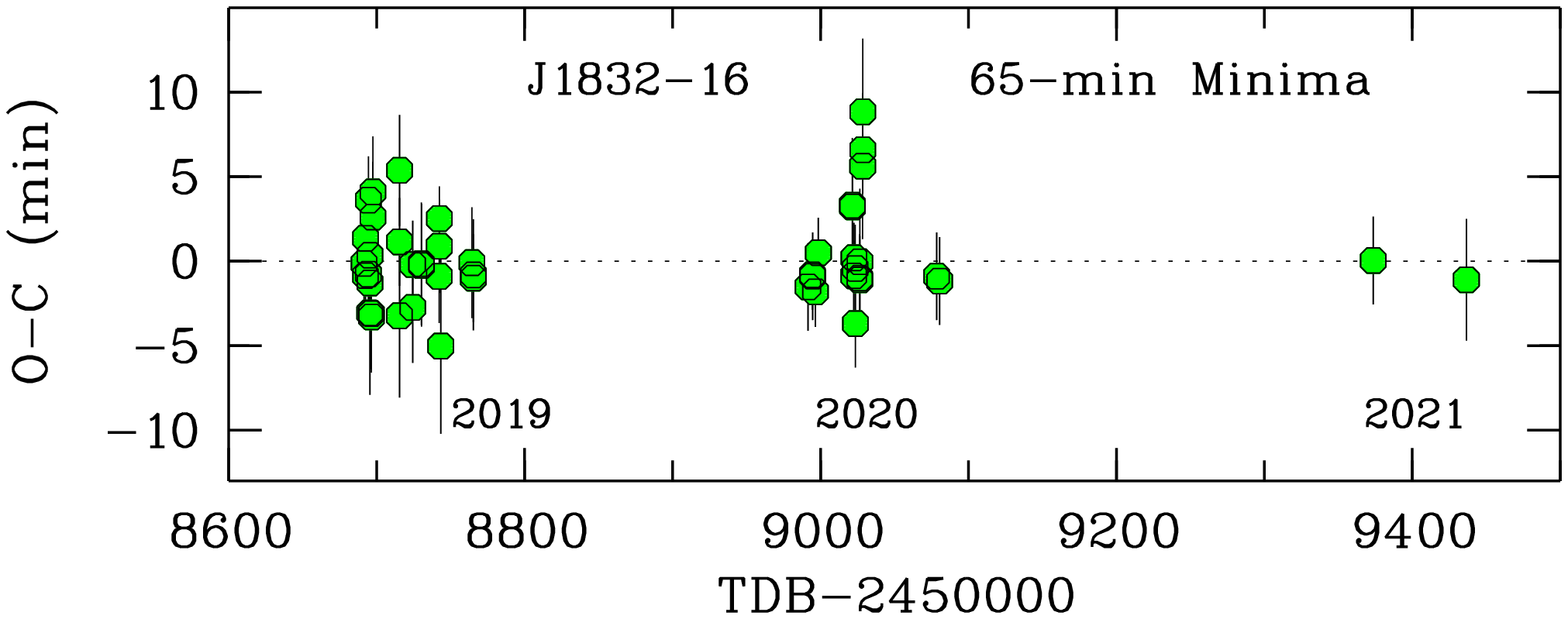}
\vspace{1.0mm}
\includegraphics[height=90mm,angle=270,clip]{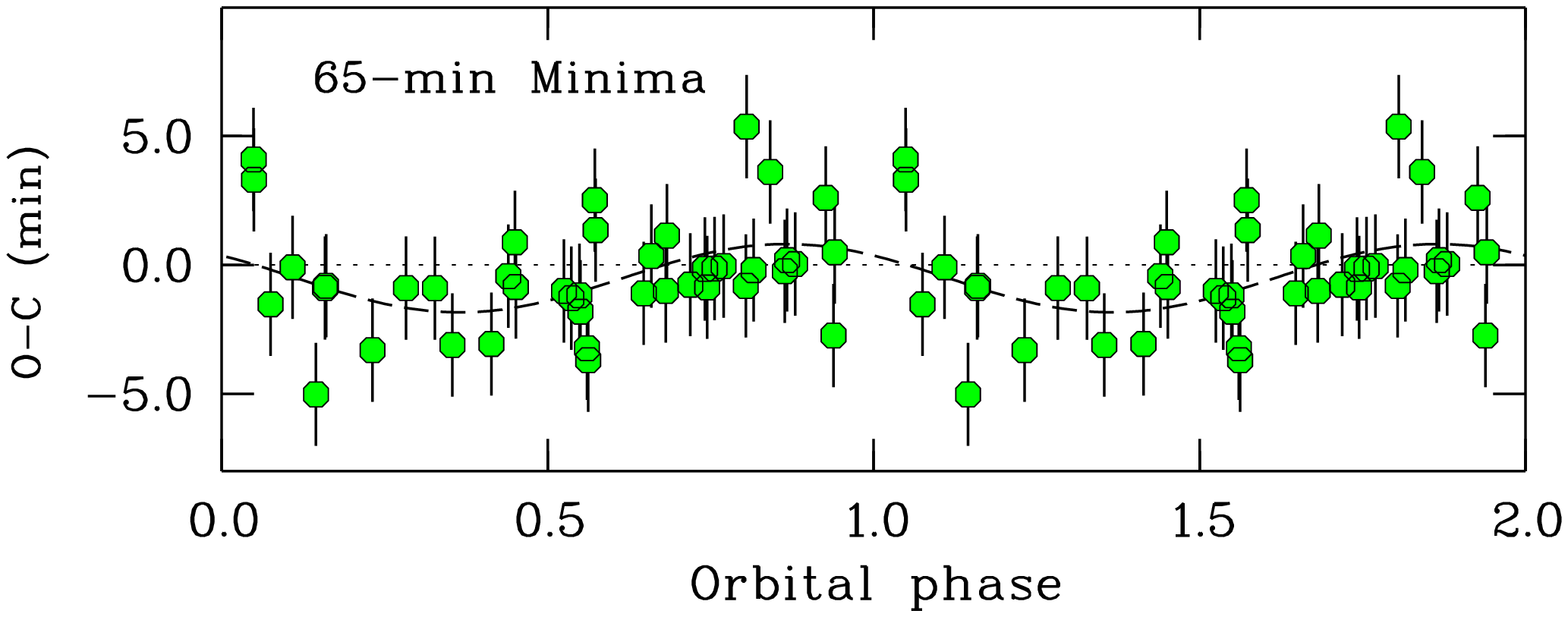}
\caption[chart]{\emph{Top: } $O-C$ diagram of the mid-eclipse
    times relative to the linear ephemeris of Eq.~\ref{eq:orb1}.  The
    difference between the quadratic and linear ephemerides of
    Eqs.~\ref{eq:orb2} and \ref{eq:orb1} is shown as the dashed
    curve. \emph{Second from top: } FWHM of the eclipse vs. pulse
    phase from Eq.~\ref{eq:spin}. \emph{Center: } $O-C$ values of the
    mid-eclipse times relative to the quadratic ephemeris
    vs. pulse phase. The model curve is from Eq.~\ref{eq:shift1}
    (see text).  \emph{Second from bottom: } $O-C$ diagram for the
    minimum times of the 65-minute pulsations relative to the linear
    ephemeris of Eq.~\ref{eq:spin}. \emph{Bottom: } $O-C$ values of
    the previous graph vs. orbital phase (see text).}
\label{fig:oc}
\end{figure}

We fitted the ingress into or the egress from eclipse by an analytical
model of the occultation of a circular disk of uniform surface
brightness by an opaque secondary star. For a given individual
exposure, the model value was represented by a Simpson integral of the
time-dependent model flux over the finite exposure time, which
provides an optimal description of the poorly resolved transitions.
The fit parameters for either ingress or egress were (i) the flux
before the transition, (ii) the flux after the transition, (iii) the
time, at which 50\% of the depth is reached, (iv) the duration of the
transition between the contact points, (v) the slopes of the
time-dependent flux before the transition, and (vi) the slope after
the transition. Additional parameters account, as far as possible, for
a systematic variation in the atmospheric conditions, for example, an
air-mass-dependent transmission. The mid-eclipse time was calculated
as the mean of the two times that mark the 50\% points at egress and
ingress, and the full width at half maximum (FWHM) or duration of the
eclipse as their difference. The light curves in Fig.~\ref{fig:lc}
show that the steps in flux and the slopes may differ substantially
between ingress and egress, depending on the respective phase in the
65-minute pulsation. It also illustrates the need to include the
slopes of the time-dependent fluxes into the fit. The derived
mid-eclipse times and the $O-C$ values relative to a linear ephemeris
are listed in Table~\ref{tab:orb} and the $O-C$ values are shown in
the top panel of Fig.\,\ref{fig:oc}.  The alias-free linear ephemeris
for all data is
\begin{equation}
T_\mathrm{ecl,1} ~= ~\mathrm{TDB}~2458691.53666(3)+0.36948844(2)\,E.~~~
\label{eq:orb1}
\end{equation}
with $\chi^2 = 387.8$ for 25\,dof (degrees of freedom) and 1$\sigma$
errors in the last digits quoted in brackets. Fitting a quadratic
ephemeris to all data gives
\begin{eqnarray}
T_\mathrm{ecl,2} & = & \mathrm{TDB}~2458691.53630(2)+0.36948972(1)\,E\nonumber\\
             &&  \hspace{35mm} -\,\ten{6.5(1)}{-10}\,E^2 
\label{eq:orb2}
,\end{eqnarray}
with $\chi^2\!=\!97.2$ for 24\,dof, which is still not good. The
difference between the two fits of Eqs.~\ref{eq:orb2} and
\ref{eq:orb1}, $T_\mathrm{ecl,2} - T_\mathrm{ecl,1}$, is shown as the
dashed curve in the top panel of Fig.\,\ref{fig:oc}. From
Eq.~\ref{eq:orb2}, the period derivative is $\dot
P_\mathrm{orb}\!=\!-\ten{3.5}{-9}$\,s\,s$^{-1}$. Taken at face value,
this implies a current
period decrease of 0.11\,s per year and a timescale of the variation
of $\ten{3}{5}$\,yr, about two orders of magnitude faster than that
predicted by nuclear evolution \citep[e.g.,][]{kalomenietal16}.

\begin{figure}[t]
\begin{center}
\includegraphics[height=70mm,angle=270,clip]{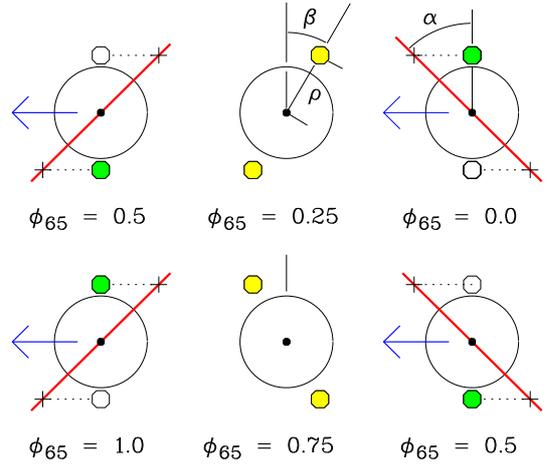}
\caption[chart]{Schematic sketch of the two-pole accreting WD passing
  the limb of the secondary star, inclined by an angle $\alpha$ to the
  vertical (red line). The emission regions are located at radial
  distance $\rho$ and displaced by an angle $\beta$ from the
  rotational axis. They are indicated by circles, color-coded for
  their relative brightness: maximal (green), intermediate (yellow),
  and minimal (open). The projected views are shown for pulse phase at
  the center of the eclipse, \phirotc$\,=\!0.25$ (top) and $0.75$
  (bottom). The pictures refer to the ingress (right), mid-eclipse
  (center), and egress (left).  The transit takes almost exactly one
  half of the observed pulse period. The blue arrows indicate the
  motion of the WD.}
\label{fig:geo}
\end{center}
\end{figure}

\subsection{Evidence for pole flipping}
\label{sec:flip}

Accounting for a period change improves the fit to the mid-eclipse
times, but with $\chi^2_\nu\!=\!4.05$, Eq.~\ref{eq:orb2} still fails
to represent the eclipse times well. We searched for a further
dependence and found that the mid-eclipse times varied
quasi-sinusoidally with the 65-minute pulse phase \phirotc\ measured at
the center of the eclipse (Fig.~\ref{fig:oc}, center panel).
%
%
A sine fit, $\Delta\,T_\mathrm{ecl}\!=\!A\,\mathrm{sin}(\varphi_\mathrm{c} -
\varphi_0) -\gamma$, gave an amplitude $A\!=\!13.6\pm1.6$\,s, a
zero-crossing at $\phi_0\!=\!0.03\pm0.02$, and
$\gamma\,=\!-0.5\pm1.1$\,s, with $\chi^2\!=\!25.9$ for 24 dof. For
brevity, we use here the phase angle $\varphi_\mathrm{c}$ at the
center of eclipse instead of $2\pi\phi_\mathrm{65,c}$. The final model
of Eq.~4 below, depicted in the center panel Fig.~\ref{fig:oc} center,
provided a small improvement over the sine fit with $\chi^2\!=\!24.0$
for 23 dof. No modulation similar to that of the mid-eclipse times is
seen in the FWHM or duration of the eclipse (Fig.~\ref{fig:oc}, second
panel from top). The weighted mean FWHM of 17 completely covered
eclipses is $1970$\,s with a standard deviation of the distribution of
2\,s.  The pulsations are discussed in Sect.~\ref{sec:osc}, their
ephemeris is provided in Eq.~\ref{eq:spin}, and its zero point is
defined as the pulse minimum.

\begin{figure*}[t]
\hspace{10mm}
\begin{minipage}[t]{102mm}
\includegraphics[height=150.0mm,angle=270,clip]{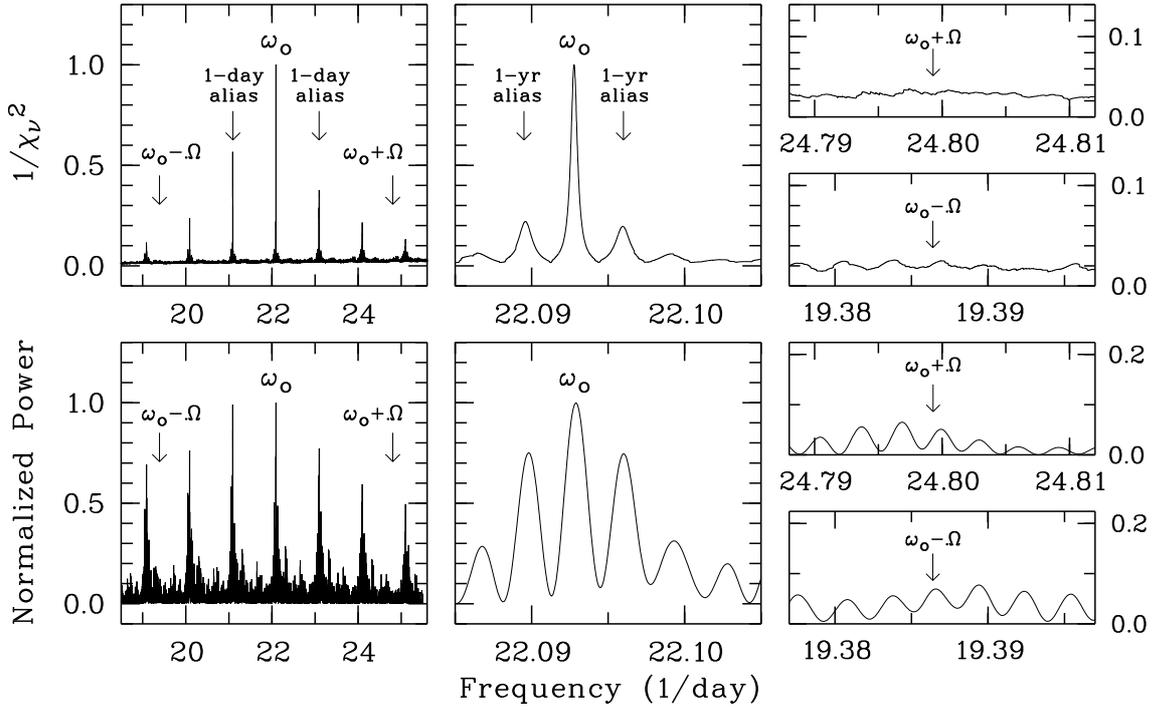}
\end{minipage}
\caption[chart]{\emph{Top: } Periodograms for the minimum times of the
  65-minute pulsations. The scale on the left also applies to the center
  panels. \emph{Bottom: } Lomb-Scargle power spectra for the high- and
  intermediate-state pulsations outside the eclipse, normalized to a
  peak value of unity. }
\label{fig:peri}

\vspace{5mm}
\end{figure*}

The shifts of the mid-eclipse times follow naturally in an IP model
with pole flipping that features two accretion columns with brightness
varying in antiphase. For simplicity, we adopted sinusoidal variations
and represented the columns as point sources that are located at the
centroid of the emission at a radial distance $\rho\!>$\,\rwd\ and an
angular offset $\beta$ from the rotational axis
(Fig.~\ref{fig:geo}). They are eclipsed by the limb of the secondary
star that is modeled as a straight edge inclined by an angle $\alpha$
against the vertical (the preferred dynamic models in
Sect.~\ref{sec:sys} have $\alpha$ = 47\degr\ or 53\degr).  Special
about \jshort\ is the near equality of the FWHM of the eclipse
(1970\,s) with one half of the observed 65-minute pulse period (1955\,s),
implying that the phase angles at ingress and egress are related by
$\varphi_\mathrm{i}\simeq\varphi_\mathrm{c}-\pi/2$ and
$\varphi_\mathrm{e}\simeq\varphi_\mathrm{c}+\pi/2$, respectively.
We assumed that the brightness of the sources near both poles varied
as $f_1\!=\!1+a_1\,\mathrm{cos}\,\varphi$ for the upper pole and
$f_2\!=\!1-a_2\,\mathrm{cos}\,\varphi$ for the lower one. With
$a_2\!>\!a_1$, the pulse maximum occurs at $\varphi\!=\!\pi$ or
\phirot$\,=\!0.5$ and is dominated by the lower pole. The phasing is
as observed (Fig.~\ref{fig:lc} and Eq.~\ref{eq:spin}) and the
preference of the lower pole is adapted from the accretion geometry of
EX Hya \citep{beuermannosborne88,rosenetal88,allanetal98}.  Both poles
display their maximal brightness difference at ingress and egress for
central phases of 0.25 and 0.75 (Fig.~\ref{fig:geo}, green = maximum,
yellow = intermediate, open = minimum). For central phases \phirotc=0
or 0.50, on the other hand, ingress and egress occur at \phirot=0.25
or 0.75 with both poles at intermediate brightness.  The dotted lines
in Fig.~\ref{fig:geo} denote the displacements from the limb and the
time shifts of ingress and egress relative to that experienced by the
center of the WD. The time shifts at both poles are formally the same
and differ only by their sign,
\begin{equation}
\Delta t = \pm(\rho/\upsilon)(\mathrm{cos}\,\beta\,\mathrm{tan}\,\alpha +
\mathrm{sin}\,\beta\,\mathrm{sin}\,\varphi),
\label{eq:shi}
\end{equation}
with $\varphi$ the current phase angle and $\upsilon$ the orbital
speed of the WD relative to the secondary star.  We estimated the
combined effect $\Delta\,T_\mathrm{ecl}$ on the mid-eclipse time by
averaging the four contributions flux-weighted with $f_1$ and
$f_2$. Similarly, we estimated the FWHM as the difference of the
weighted mean contributions at egress and ingress. Our simple model
predicts that the FWHM does not vary with $\varphi_\mathrm{c}$, but
that substantial shifts of the mid-eclipse times occur,
\begin{equation}
\Delta\,T_\mathrm{ecl}=A\,[\mathrm{cos}\,\beta\,\mathrm{tan}\,\alpha\,\mathrm{sin}\varphi_\mathrm{c} - (1/2)\mathrm{sin}\,\beta\,\mathrm{sin}(2\varphi_\mathrm{c})], \\
\label{eq:shift1}
\end{equation}
with an amplitude $A\!=\!(a_1\!+\!a_2)\rho/2\upsilon$. The observed
dependence of $\Delta T_\mathrm{ecl}$ on $\varphi_\mathrm{c}$ (or
\phirot) in the center panel of Fig.~\ref{fig:oc} is only moderately
skewed, suggesting that the first term in Eq.~4 dominates.  We fitted
this model to the observed time shifts and obtained
$\beta=21\!\pm10$\degr, an amplitude
$A\!=\!(a_1\!+\!a_2)\rho/2\upsilon\!=\!(12.2\!\pm\!1.4)$\,s for
$\alpha\!=\!50$\degr\ and $\chi^2\!=\!24.0$ for 23 dof, as quoted
already above. With these parameters, Eq.~\ref{eq:shift1} becomes
\begin{equation}
\Delta\,T_\mathrm{ecl}=0\fd00016(2)\,[\mathrm{sin}\varphi_\mathrm{c} - 0.16\,\mathrm{sin}(2\varphi_\mathrm{c})], \\
\label{eq:shift2}
\end{equation}
with an amplitude of 14\,s and
$\varphi_\mathrm{c}\!=\!2\pi\phi_\mathrm{65,c}$ defined by
Eq.~\ref{eq:spin}, below. For $\beta>40$\degr, the model curve becomes
excessively skewed and develops double humps. The amplitudes
$a_1\!\simeq\!0.2$ and $a_2\!\simeq\!0.8$ fit the observed mean 65-min
light curve in Fig.~\ref{fig:lc}. The orbital speed is
$\upsilon\!\simeq\!\ten{3.1}{7}$\,\cms\ for a binary separation
$a\!\simeq\!\ten{1.6}{11}$\,cm (Sect.~\ref{sec:sys}) and the
characteristic size of the columns is a plausible
$\rho\!\simeq\!10^9$\,cm, similar to the value estimated by
\citet{siegeletal89} for EX Hya.  Correcting the original mid-eclipse
times for the shifts defined by Eq.~4, linear ephemerides for the
individual years gave period differences for 2019 and 2021 relative to
2020 of $+(94\pm39)$\,ms and $-(95\pm52)$\,ms, respectively. A
complete orbital ephemeris, given by the addition of Eqs.~\ref{eq:orb2}
and~\ref{eq:shift2}, requires knowledge of the pulse phase.
  
The subsequent eclipses of both poles are expected to lead to a
phase-dependent structure of the ingress and egress light curves that
stayed unresolved in our data, but may provide insight into the
accretion geometry, if measured with higher time resolution. We
emphasize that such a study was not previously possible, because no
deeply eclipsing stream-fed IP was known.

Our simple model of Eq.~\ref{eq:shift1} does not account for more
complex geometrical effects and radiative transfer in the emission
regions. Consequently, it does not account for variations in the pulse
profile as a function of orbital phase.  Such variations are
predicted, however, in the more elaborate model of
\citet{ferrariowickramasinghe99} for a stream-fed IP accreting
alternately at two opposite poles. The model light curves presented in
their Fig.~4, feature double-peaked pulses of the line emission and
the optical continuum from the extended accretion columns, while the
X-ray pulses are single-peaked, because of the two accretion spots on
the WD only one is visible at any given instant. The optical pulses
display phase shifts over the orbital period and a variation in the
separation of the double peaks, including a short single-peaked
interval. We measured the minimum times of the continuum flux from
Fig.\,4 in \citet{ferrariowickramasinghe99} and found that they
display a quasi-sinusoidal variation over the orbit. Such a variation
is also observed in our data and shown in the bottom panel of
Fig.~\ref{fig:oc}. The amplitude of a fitted sinusoid is
$-1.5\!\pm\!0.5$\,min (where we have excluded the three outlying
minimum times of 2020 in the second panel from the bottom, which show
no phase preference).
We searched for an orbital variation in the pulse shape in our WL data
by sorting all individual WL measurements into an image of $15\times
15$ orbital and pulsational phase bins (not shown). In that image, the
shifting pulse minimum is even more clearly visible than in the bottom
panel of Fig.~\ref{fig:oc}. In addition, there is an indication that
the separation of the two peaks varies over the orbit including what
may be a short interval with a central single peak near orbital phase
$\phi\!=\!0.5$. Given improved observations, dedicated modeling may be
rewarding. Presently, our findings strongly suggest that pole-flipping
occurs in \jshort, supporting its classification as a stream-fed IP.

Finally, we caution against prematurely identifying the $O-C$ variation
in the top panel of Fig.~\ref{fig:oc} and the quoted period
differences between the individual years as real variations in the
orbital period before the dynamics of the system is fully understood.
Orbital period changes in close binaries, mostly detached ones, are
ubiquitous and not well understood. Neither the concept of a third
body in the system nor varieties of the Applegate mechanism
\citep{applegate92,voelschowetal18,lanza20} have so far led to a
generally accepted explanation of the observed variations.

\subsection{Pulse period and ephemeris}
\label{sec:osc}

We employed two methods to determine the period of the pulsation:
(i)~calculating a Lomb-Scargle power spectrum in the MIDAS-TSA
context\footnote{ESO MIDAS Manual, Part B, Sect. 12.4, MIDAS utilities
  for Time Series Analysis} and (ii)~subjecting the minimum times of
the pulsations to a period search. The former uses the entire
information from the rather complex pulse profile, the latter only the
reduced information of a well-defined feature with a lower internal
scatter. The times of the minima were determined graphically by the
bisected-chord technique, marking the center between descent into and
ascent from minimum at different flux levels and extrapolating these
times to the flux at the minimum. The standard deviation of the
distribution of the $O-C$ values of the 47 minima of Table~\ref{tab:spin}
around a linear fit is 2.9\,min. For comparison, locating an
individual maximum may prove complicated and the uncertainty may
exceed 10 min or more. Consequently, the Lomb-Scargle periodogram is
expected to be less efficient in measuring the pulse period, but is
free of the necessarily somewhat subjective procedure of defining the
minima.

The Lomb-Scargle power spectrum for the 2019--2021 high- and
intermediate-state observations outside the eclipse, is displayed in
the lower left panel of Fig.\,\ref{fig:peri}, normalized to a peak
value of unity. Maximum power is attained at an observed frequency of
the pulsation $\omega_\mathrm{o}\!=\!22.09$\,day$^{-1}$, closely
followed by the 1-day alias at 21.09\,day$^{-1}$. The center bottom
panel shows that at higher resolution power appears also at the 1-year
alias frequencies separated from the best value by 0.003\,day$^{-1}$,
reflecting the separation of the 2019 and 2020 observing periods by
about 320 days. The two right-hand panels demonstrate that there is no
power at the side bands $\omega_\mathrm{o}\!\pm\!\Omega$, where
$\Omega\!=\!2.70644(1)$\,day$^{-1}$ is the orbital frequency. The same
holds for the side band at $\omega_\mathrm{o}\!-\!2\Omega$
\citep{warner86} and is true also for the corresponding side bands
that belong to the alias at $\omega_\mathrm{o}\,=\!21.09$\,day$^{-1}$.
Not surprisingly, in view of the double peaks, there is significant
power at the second harmonics of observed pulse frequency
$\omega_\mathrm{o}$ and its alias frequencies (not shown).
Gratifyingly, the periodogram yields the same frequencies as the
Lomb-Scargle power spectrum, but more clearly prefers
$\omega_\mathrm{o}\!=\!22.09$\,day$^{-1}$ with a nominal reduced
$\chi^2_{\nu}$ of unity over 21.09\,day$^{-1}$ with
$\chi^2_{\nu}\!=\!1.77$.  Again, no signal is detected at any of the
side bands of $\omega_\mathrm{o}$. Its best value is
$\omega_\mathrm{o}\!=\!22.09281(4)$\,day$^{-1}$. In case of the 1-day
alias, the fit selects also a different of the 1-year alias frequencies
and the best value would become $\omega\!=\!21.090$\,day$^{-1}$.
We consider $\omega_\mathrm{o}\!=\!22.09281$\,day$^{-1}$ as the most
probable choice. The analysis of the 47 minimum times in
Table~\ref{tab:spin} yielded the linear ephemeris
\begin{equation}
T_\mathrm{min}\!=\!\mathrm{TDB}~2458691.4462(4) + 0.04526359(7)\,E.~~~
\label{eq:spin}
\end{equation}
The best period is $P_\mathrm{o}\!=\!3910.774(6)$\,s$\,=\!65.1796(1)$\,min
and the $O-C$ diagram is shown in the second panel from below in
Fig.\,\ref{fig:oc}. Fitting the minimum times of 2019 and 2020
individually gave periods of 65.1798(10)\,min and 65.1806(14) min,
respectively. There is no evidence for a period variation so far
and we accept the pulsation in \jshort\ as a stable coherent signal.
We cannot presently identify the observed period uniquely with either
the spin period or the synodic (beat) period of the WD.  For the two
cases, the spin period is either $P_\mathrm{spin}\!=\!3910.775(7)$\,s
or 3483.976(7)\,s and the ratio of spin versus orbital periods is
\pspo$\,=\!0.1225$ or 0.1091, respectively.

\begin{figure*}[t]
\includegraphics[height=105.0mm,angle=270,clip]{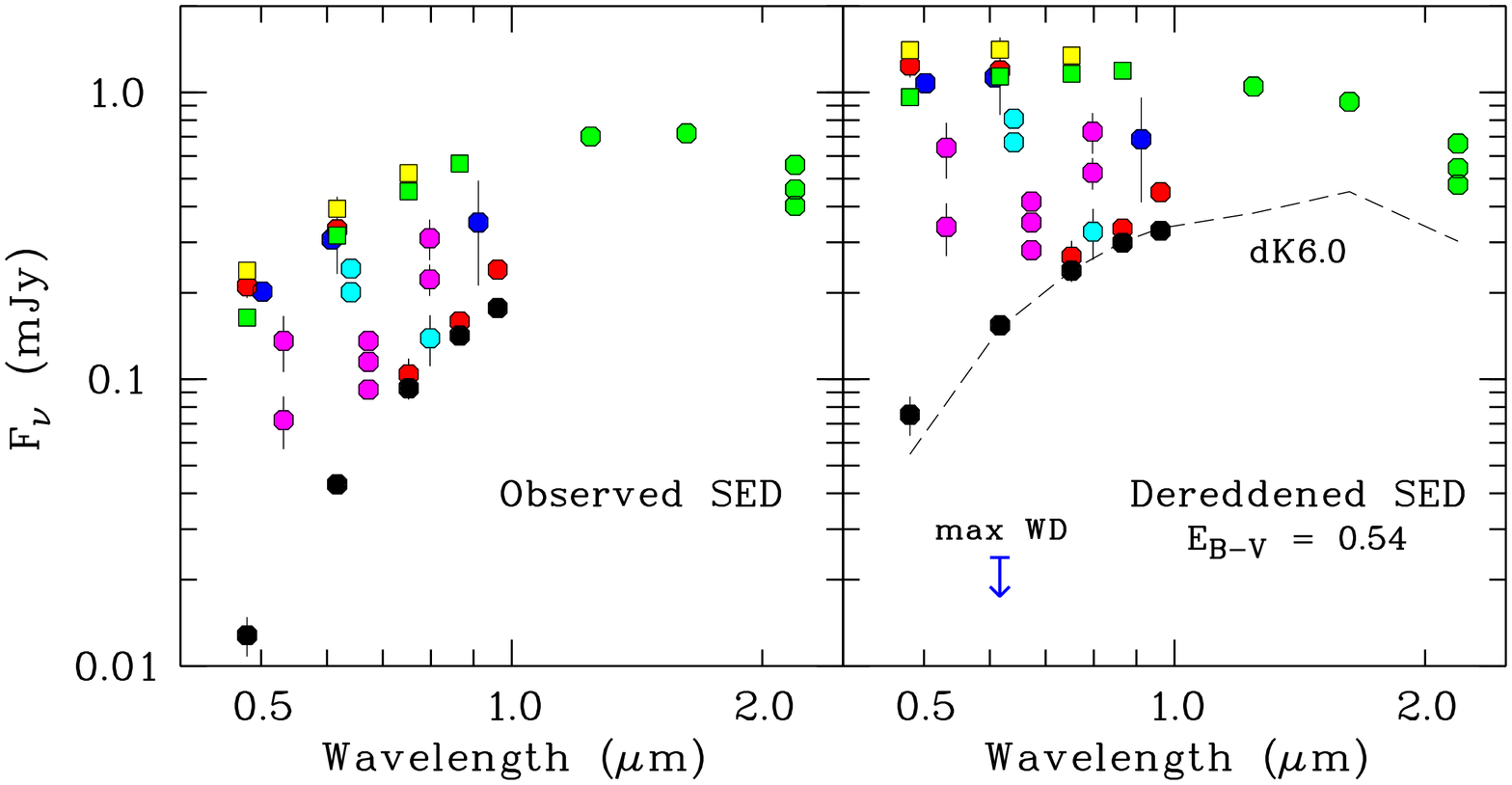}
\hfill  
\includegraphics[height=74.0mm,angle=270,clip]{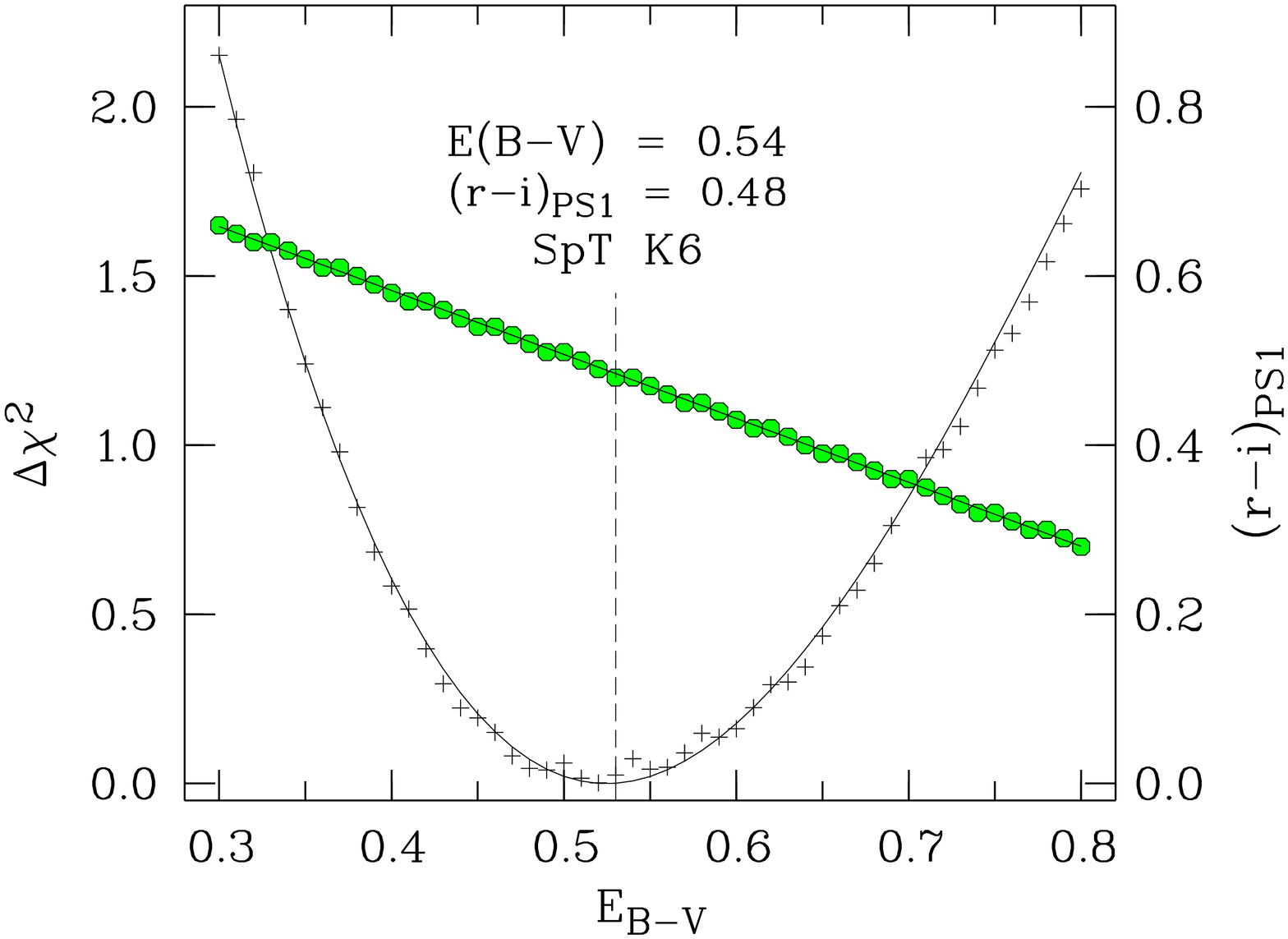}
\caption[chart]{\emph{Left: } Observed SED of \jshort. Black dots
  refer to the minimum fluxes in the eclipse and represent our best
  estimate of the SED of the secondary star. Green and yellow squares
  indicate the SED outside eclipse measured simultaneously with the
  MuSCAT3 photometer and the FTN on 9 May and 15 April
  2021, respectively. Colored dots indicate nonsimultaneous
  out-of-eclipse fluxes (see text). \emph{Center: } De-reddened SED
  based on the best-fit $E_\mathrm{B-V}\!=\!0.54$.  \emph{Right: }
  $\Delta \chi^2$ vs. $E_\mathrm{B-V}$ (plus signs, left scale) and the
  associated $(r-i)_\mathrm{P1}$ (green dots, right scale).}
\label{fig:sed}
\end{figure*}

\section{Multicolor spectral energy distribution}
\subsection{Secondary star} 
\label{sec:sec1}

In this section we consider the $grizy$ eclipse measurements of
\jshort\ performed with the Faulkes North and South telescopes and
derive the spectral energy distribution (SED) of the secondary
star. We measured the spectral fluxes of the target in the central
part of the eclipse, using the method described in the last paragraph
of Sect.~\ref{sec:obs}. These fluxes varied by about 10\% in $rizy$
and more in $g$. We adopted generally the lowest of the
photometrically well defined fluxes as representative of the SED of
the secondary star, list these in Table~\ref{tab:eclmag}, and show
them as black solid circles in the left panel of Fig.~\ref{fig:sed}.

\begin{table}[b]
\begin{flushleft}
\caption{Energy fluxes and Pan-STARRS1 AB magnitudes of \jshort\ at the
  center of the eclipse.}
\vspace{-2mm}
\begin{tabular}{@{\hspace{1.0mm}}c@{\hspace{4.0mm}}c@{\hspace{4.0mm}}l@{\hspace{3.0mm}}c@{\hspace{3.0mm}}c@{\hspace{3.0mm}}c}\\[-1ex]
  \hline\hline                                                          \\[-1.5ex]
  Band &Observed  & Flux C3  & Flux Target& Target &  $A_\mathrm{\lambda}$\\
  (P1)&Flux ratio&  (mJy)   & ($\mu$Jy)      & (AB mag)     & (mag)       \\[0.5ex]
  \hline                                                            \\[-1ex]                                                   
   $g$   & 0.021\,(2) & 0.612\,(2) & ~~13\,(1) & \hspace{1.5mm}21.12\,(10)  & 3.560    \\       
   $r$   & 0.041\,(1) & 1.04\,(1)  & ~~43\,(1) & 19.83\,(3)   & 2.567    \\
   $i$   & 0.071\,(2) & 1.31\,(1)  & ~~93\,(3) & 18.98\,(3)   & 1.899    \\
   $z$   & 0.092\,(3) & 1.54\,(1)  & 142\,(5)  & 18.52\,(4)   & 1.498    \\
   $y$   & 0.111\,(5) & 1.59\,(1)  & 177\,(8)  & 18.28\,(5)   & 1.249    \\[1.0ex]
 \hline\\
\end{tabular}\\[-1.0ex]
\label{tab:eclmag}
\end{flushleft}
\vspace{-2mm}
\end{table}

Fitting the observed SED of the secondary star requires de-reddening
it. The reddening at the position of \jshort\ was studied by
\citet{lallementetal18} within the framework of their 3D-model of the
galactic extinction and is in the range of \ebmv\ = 0.3-0.7 for
distances between 900 and 2100\,pc.  The geometric distances $d$
quoted in the Gaia data releases DR2 and EDR3 have lower and upper
error bounds of $941-2696$\,pc and $1254-2847$\,pc, respectively
\citep{bailerjonesetal18,bailerjonesetal21}. Models of \jshort\ that
include \ebmv\ and the distance $d$ as independent parameters are
presented in Sect.~\ref{sec:sys}.  Here, we vary \ebmv, fitting the
de-reddened relative $rizy$ fluxes with a stellar SED
\citep{tonryetal12,coveyetal07,pecautmamajek13} to obtain the best-fit
color \ri\ for a given \ebmv\ and thereby the spectral type. For
consistency with the observations, all colors were transformed into
the Pan-STARRS P1 system.  Since the $g$-band flux exceeded the
predicted stellar flux slightly, but systematically (see
Sect.~\ref{sec:disk} and Fig.~\ref{fig:ecl}), we restricted the fit to
the $rizy$ bands. The fit gave an overall~$\chi^2$ minimum for
\ebmv$\,=\!0.54$ and \rip$\,=\!0.48$, with 1$\sigma$ ranges ($\Delta
\chi^2\!=\!+1$) of the two quantities of $0.37-0.71$ and $0.35-0.61$,
respectively. The errors are rather large because the reddening path
in the color-color diagram runs more or less parallel to the stellar
locus. The green dots in the right panel of Fig.~\ref{fig:sed} show
the relation between \rip\ and \ebmv,\\
\begin{equation}
(r-i)_\mathrm{P1}\,=\!0.8855\!-\!0.7563\,E_\mathrm{B-V}.
\label{eq:rieb}
\end{equation}
The derived range of \ebmv\ agrees with the above quoted result of
\citet{lallementetal18} at the position of \jshort\ for distances
between 1 and 2\,kpc. The SED of the secondary star of
\jshort\ de-reddened with \ebmv$\,=\!0.54$ is displayed in the center
panel of Fig.\,\ref{fig:sed} (black solid circles). The implied
spectral type of the secondary star depends on the employed
algorithm. \citet{coveyetal07}, for instance, used ``the Hammer''
\citep{kesselietal20}, which relates \rip$\,=\!0.48$ and its
confidence range to K6.0 and K3.9 to K7.5. The spectral types assigned
to the same \rip\ in the system of
\citet{pecautmamajek13}\footnote{http://www.pas.rochester.edu/$\sim$emamajek/EEM\_dwarf\_UBVIJHK\_
  colors\_Teff.txt, version 2019.3.22.} are about half a subclass later.

\subsection{Limit on the photospheric flux of the WD}
\label{sec:wd}

The minute excess of the out-of-eclipse flux over the eclipse flux in
the low-state observation of 4 August 2019 (left panel of
Fig.~\ref{fig:ecl}, blue dots) allowed us to set an upper limit on the
WL photospheric flux of the WD. The excess of the flux at orbital
phases $\phi_\mathrm{orb}\!=\!-0.065$ to $-0.032$ and
$\phi_\mathrm{orb}\!=\!0.032$ to $0.038$ over the eclipse flux between
$\phi_\mathrm{orb}\,=\!-0.0295$ and $+0.0295$ corresponds to a WL AB
magnitude of $w\!=\!21.84\,\pm\,0.20$, which qualifies approximately
also as an $r$ magnitude (see Sect.~\ref{sec:obs}). A 2$\sigma$ upper
limit to the WD is $r\!>\!21.44$. On 17, 20, and 27 July 2021,
\jshort\ was found in a low state again and observed in the $g$-band
through the eclipse. In all three nights, the flux difference between
outside and inside the eclipse was consistent with zero and a combined
2$\sigma$ upper limit for the WD is $g\!>\!21.69$.

\subsection{Overall spectral energy distribution of \jshort}
\label{sec:sec2}

In the left panel of Fig.~\ref{fig:sed} we provide an overview of the
SED of \jshort\ outside eclipse as well, based on a mixture of
simultaneous and nonsimultaneous flux measurements of our own and
from public catalogs accessed mainly via the Vizier SED
tool\footnote{Provided by the Centre de Donn\'ees astronomiques de
  Strasbourg http://vizier.unistra.fr/vizier/sed/}. The external
sources included the NOMAD catalog \citep{zachariasetal05}, the
SkyMapper catalog \citep{wolfetal19}, the Pan-STARRS catalog
\citep{chambersetal17}, and the Gaia catalog \citep{gaia21}.  No entry
was found in the GALEX catalog \citep {bianchietal17}. The object is
too faint for 2MASS \citep{skrutskieetal06}, but is listed in the
UKIRT Galactic plane survey \citep{lucasetal08}. The dynamic range of
the observed fluxes increases toward short wavelengths. We take the
upper envelope to the observed spectral flux $F_\mathrm{\nu }$ as
representative of the high state of the system and obtain the
accretion-induced spectral flux by correcting it for the secondary
star. The optical component of the accretion-induced flux is obtained
by de-reddening it with the selected \ebmv\ and integrating it from the
Balmer edge into the infrared. The center panel shows that the
spectral flux de-reddened with the same \ebmv$\,=\!0.54$ as the eclipse
fluxes is slowly rising toward short wavelengths. For this choice of
\ebmv, the optical component of the accretion-induced flux is
$F_\mathrm{opt}\!=\!\ten{7.4}{-12}$\,\ergs. In the models of
Sect.~\ref{sec:sys}, we use $F_\mathrm{opt}$ de-reddened with the free
parameter \ebmv. Well studied IPs, as \exhya\ \citep[][their
  Fig.\,3]{eisenbartetal02} or \vsgr\ \citep[][their
  Fig.\,2]{beuermannetal04} have SEDs that extend to the Lyman edge,
with an integrated de-reddened ultraviolet flux that exceeds the
optical flux by a factor of about two to three and considering the
X-ray regime, the total accretion induced flux can exceed the optical
flux by a still higher factor.
\jshort\ was not detected in the few days of coverage in the ROSAT All
Sky Survey \citep{bolleretal16}, which does not argue against
\jshort\ being an X-ray source, considering its frequent low states,
the sizable extinction, and the fact that there is no pointed ROSAT
observation of the region. \jshort\ is not an INTEGRAL hard X-ray source
\citep{krivonosetal17}\footnote{https://cdsarc.unistra.fr/viz-bin/cat/J/MNRAS/470/512}
and searching the NASA HEASARC archive\footnote{
  https://heasarc.gsfc.nasa.gov/cgi-bin/W3Browse/w3browse.pl} at the
position of \jshort, gave no entry for any earlier X-ray mission. It
is not an obvious bright source in the preliminary eROSITA All Sky Map
\citep{predehletal21}\footnote{https://www.mpe.mpg.de/7463606/news20200619}
with its $15\arcmin\times 15\arcmin$ pixels. It may still have been
detected in the survey, of which a source list is not yet
available. We return to the overall SED in the context of estimating
the accretion rate (Sect.~\ref{sec:aa}).

\section{System parameters}
\label{sec:sys}

\subsection{Photometric limit on $M_2$}
\label{sec:sysphot}

\begin{figure*}[t]
\includegraphics[height=59.0mm,angle=270,clip]{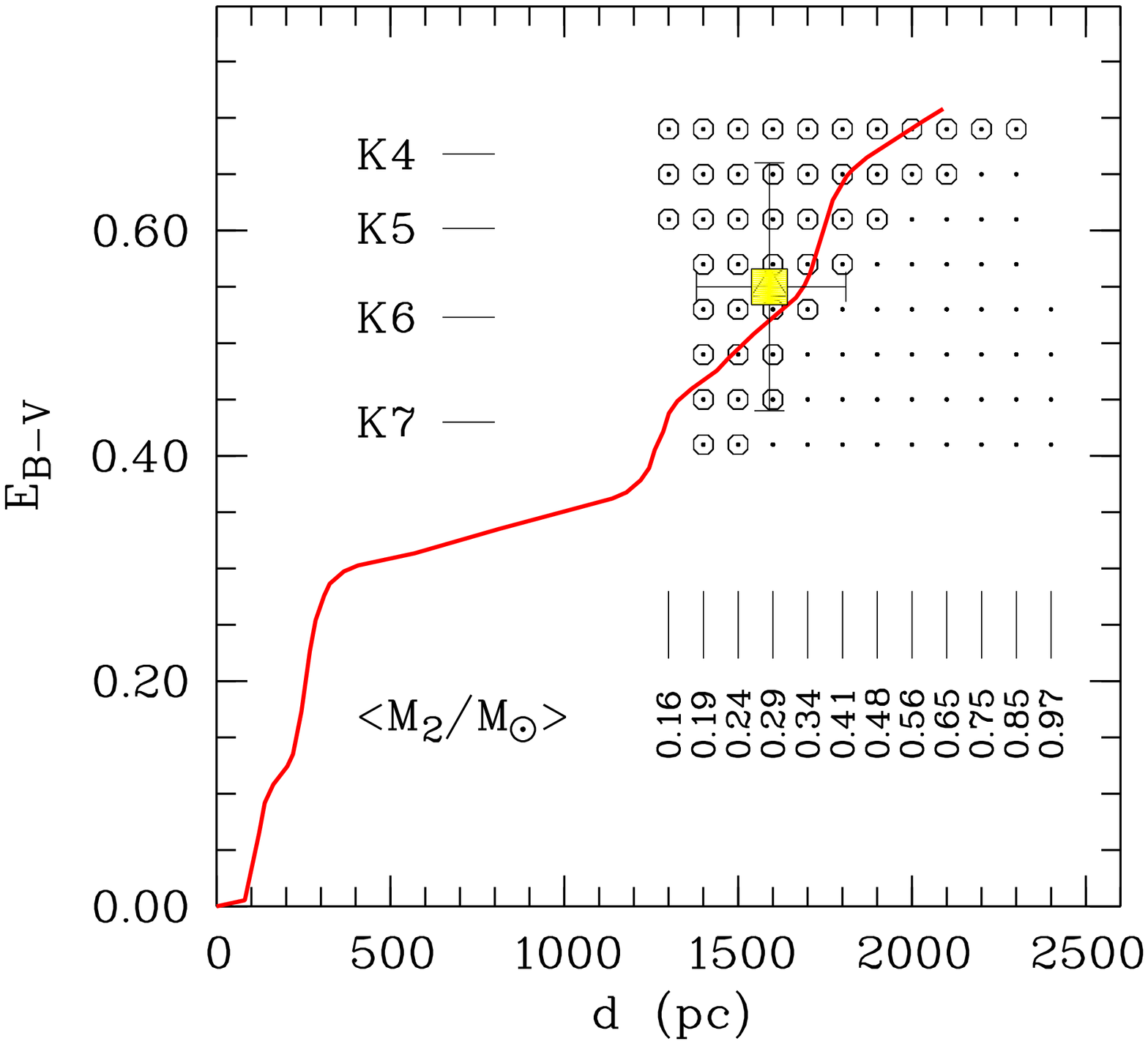}
\hspace{2mm}
\includegraphics[height=59.00mm,angle=270,clip]{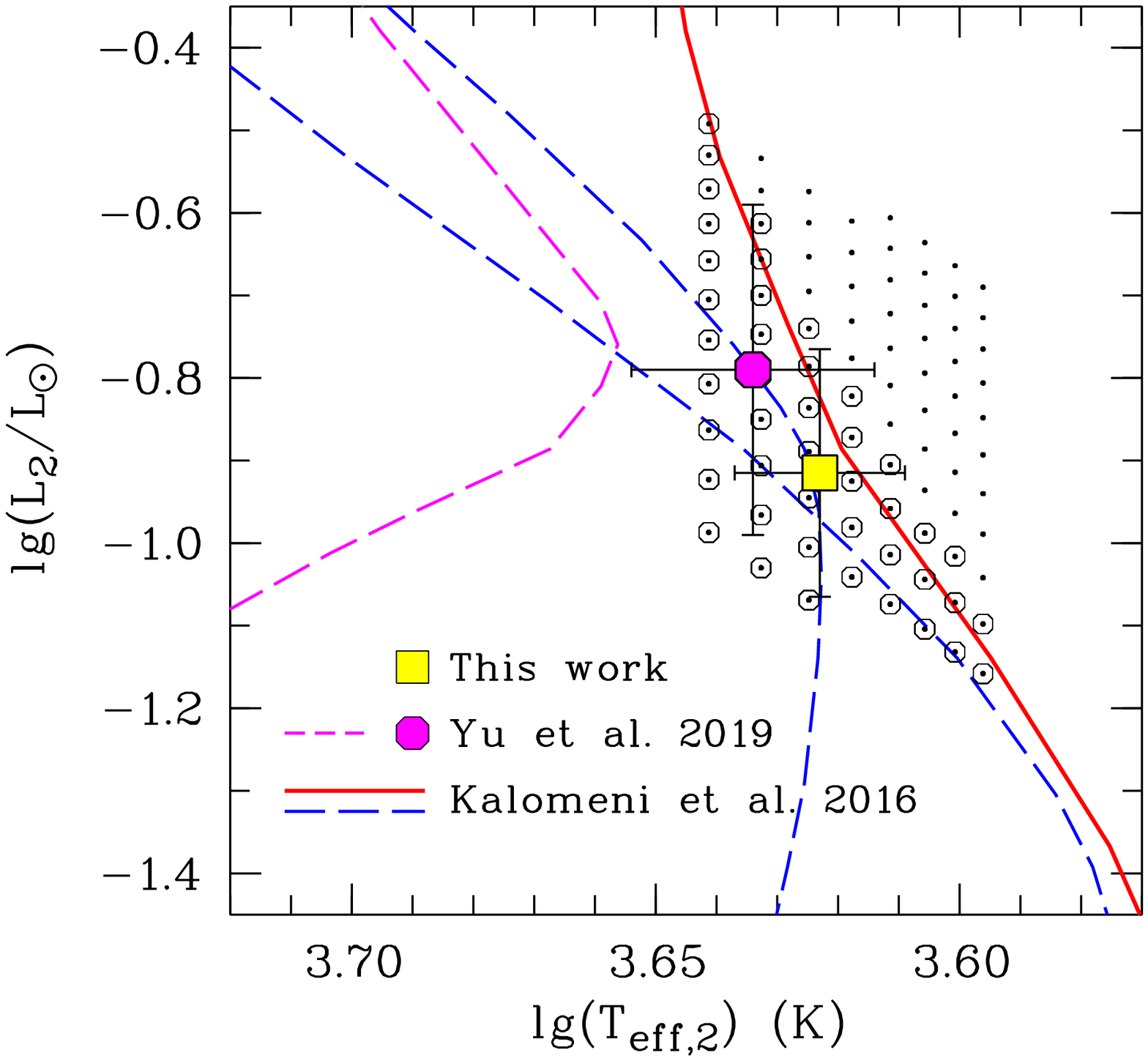}
\hspace {2mm}
\includegraphics[height=59.0mm,angle=270,clip]{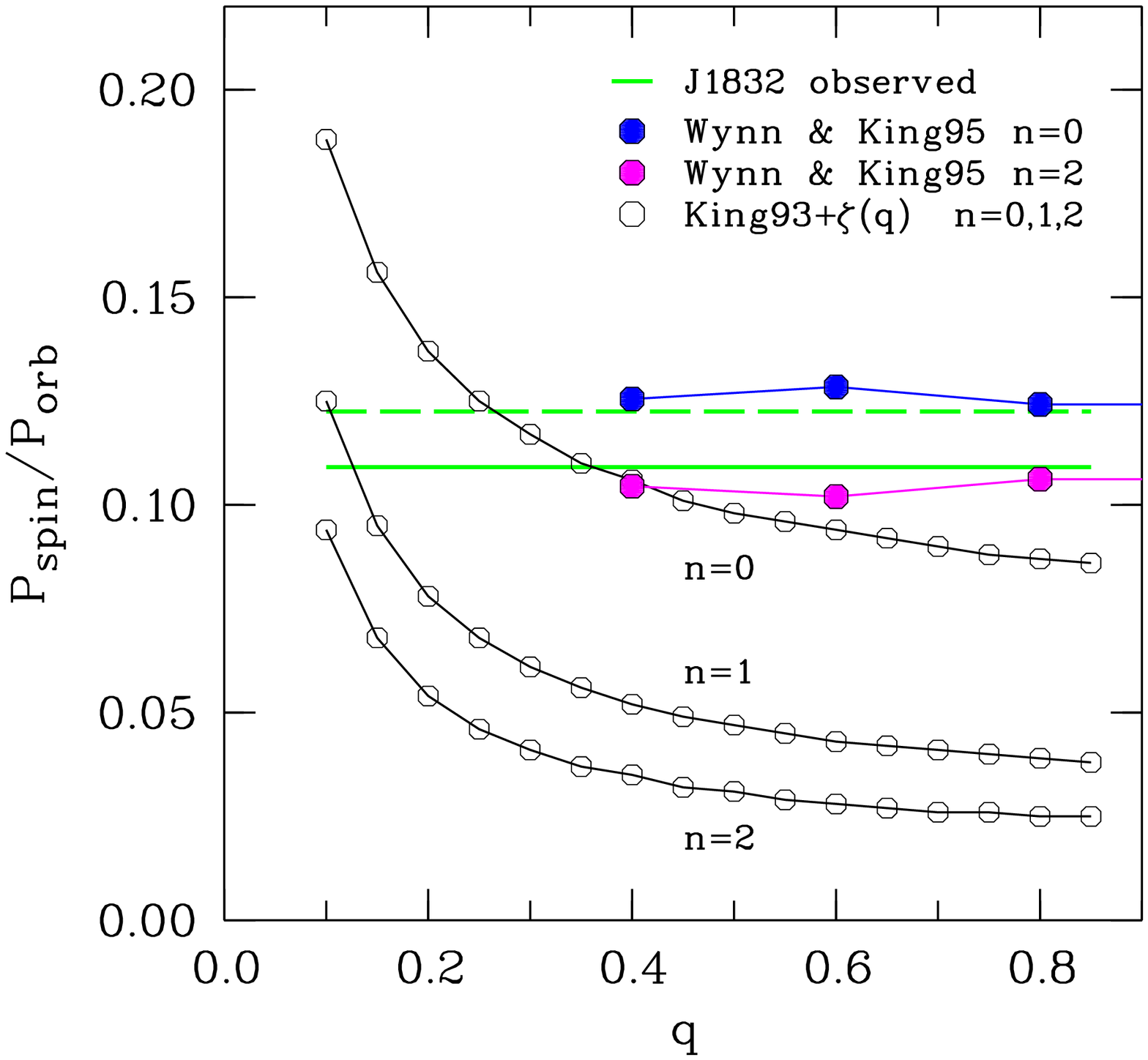}
\caption[chart]{\emph{Left: } Confidence range for models of the
  secondary star permitted by the photometric SED in the eclipse of
  \jshort\ (dots and encircled dots). All models have distances within
  the error bounds of Gaia EDR3 (footnote 3).  The red
  curve represents the best estimate of the extinction along the line
  of sight to \jshort\ \citep{lallementetal18}. \emph{Center: }
  Position of the secondary star of \jshort\ in its HR diagram. The
  dots indicate all photometrically accepted models.  The encircled
  dots denote here and in the left panel the subsample with luminosity
  $L_2$ not exceeding the limit set by the BPS of
  \citet{kalomenietal16} (red curve).  No model should be found above
  and to the right of the red curve (see text). The yellow-filled
  square in both panels marks the mean position of this subsample. The
  dashed blue curves from Kalomeni et al. (2016) indicate two
  evolutionary tracks along which \jshort\ may evolve. Also shown are
  the position of the 9~h binary \kic\ of \citet{yuetal19} (magenta
  dot) and its suggested evolutionary track that leads to the
  graveyard of ultracompact binaries (magenta dashed
  curve). \emph{Right: } period ratio \pspo\ vs. mass ratio
  $q$. The horizontal green lines indicate the observed period ratio
  of \jshort. The open and filled circles denote the theoretical
  period ratios from the analytical theory of K93 and the
  numerical calculations of WK95.}
\label{fig:sys}
\end{figure*}

In the absence of spectroscopy, the information on the system
parameters is limited. Some important information, however, can be
drawn from the measured $i$-band magnitude of the secondary star. In
particular, it allows $M_2$ to be calculated as a function of the
distance, $d$, with a minor dependence on \ebmv.  The quantities are
related by the surface brightness
$S_\mathrm{i}\!=\!M_\mathrm{i}\,+\,5\,\mathrm{lg}(R_2/R_{\odot})$,
where $M_\mathrm{i}$ is the absolute magnitude in the $i$-band and
$R_2$ the stellar radius. Using Mamajek's table of the stellar locus
(footnote 5) with colors transformed to the Pan-STARRS\,1 (P1) system,
we expressed the surface brightness for the restricted range of
\rip\,=\,0.3\,-\,0.8 as\\[-1ex]
\begin{equation}
  S_\mathrm{i} = 5.151-1.867\,(r-i)_\mathrm{P1}.
\label{eq:siri}
\end{equation}
With $i-M_\mathrm{i}\,=\!5\,\mathrm{lg}\,d\!-\!5$ and $d$ in pc, $R_2$
in solar units becomes
\begin{equation}
  \mathrm{lg}(R_2/R_\odot)\,=\!(S_\mathrm{i}-i+A_\mathrm{i}E_\mathrm{B-V})/5+\mathrm{lg}\,d-1.
\label{eq:r2}
\end{equation}
Since \jshort\ is accreting, the secondary star must fill its Roche
lobe and is necessarily bloated if of low mass. We adopted the model
radii of \citet{baraffeetal15} for main-sequence stars of solar
composition at an age of 5\,Gyr and approximated them by a power law
$R_\mathrm{BHAC}/R_\odot=a(M_2/M_\odot)^b$ with $a=0.93$ and $b=0.96$
for the mass range of $0.15-0.8$\,\msun. Secondary stars in normal CVs
are known to exceed these radii by a few percent \citep[see also
  Beuermann et al. 2017]{kniggeetal11}. We expressed the stellar radii
in our models as $R_2\!=\!f_1f_2f_3R_\mathrm{BHAC}$, with
$f_1\!=\!1.020$ and $f2\!=\!1.045$ accounting for starspots and the
effect of rotation and Roche geometry, respectively
\citep{kniggeetal11}. The factor $f_3$ describes any additional
bloating required to make the secondary star fill its lobe.
We used Paczynski's approximation $R_\mathrm{2,R}\,=\!0.462
(q/(1+q)^{1/3}a$ for the mean volume filling Roche radius, which has
the advantage that with the binary separation $a$ eliminated with help
of Kepler's third law, one obtains an expression that is independent
of $q$ and $M_1$
\begin{equation}
  M_2/M_\odot\,=\!(R_2/R_\odot)^3(2\pi/P)^2(R_\odot^3/(0.462^3\mathrm{G}M_\odot)).
\label{eq:m2}
\end{equation}
Replacing $R_2/R_\odot$ with Eqs.~\ref{eq:r2}, \ref{eq:siri}, and
\ref{eq:rieb}, $M_2$ can be calculated for given $d$ and \ebmv.  For
instance, for a distance of 1600\,pc and the best-fitting extinction
of 0.54 mag, $M_2\!=\!0.29$\,\msun. To fill its Roche-lobe, this star
would be bloated by a factor $f_3\!=\!2.27$ over its main-sequence
radius.  An unevolved secondary star of 1.0\,\msun\ as expected for a
standard CV at $P\!\simeq\!9$\,hr would require a distance of
2500\,pc, near the upper end of the confidence range of the Gaia EDR3,
$d\!=\!1254-2847$\,pc \citep{bailerjonesetal21}.

\begin{table}[bth]
\begin{flushleft}
\caption{Mean values and ranges of parameters of all photometrically
  accepted models (Cols. 3-5) and those in spin equilibrium with
  \ps$\,=\!3484$\,s (Cols. 6-8; see text for explanation).}
\vspace{-4mm}  
\begin{tabular}{@{\hspace{0.0mm}}l@{\hspace{5.0mm}}c@{\hspace{1.0mm}}c@{\hspace{1.0mm}}c@{\hspace{1.0mm}}c@{\hspace{1.0mm}}l@{\hspace{3.0mm}}c@{\hspace{1.0mm}}c}\\[-1ex]
\hline\hline                                                                                                         \\[-1.5ex]
~\#~~~Quantity    & \multicolumn{5}{c}{\hspace{-7mm}Photometrically accepted} & \multicolumn{2}{c}{\hspace{-2mm}$L_2$-limited}      \\
                         & Mean & Sigma & \multicolumn{3}{c}{\hspace{0mm} Range}~~ & Mean  & Sigma \\[0.5ex]
\hline                                                                                                  \\[-1ex]                                                   
~1~~~$d$ (pc)                              & 1755  & 284   & 1270  & $-$ & 2500   & 1596   & 208\\           
~2~~~\ebmv\ (mag)                          & 0.54  & 0.10  & 0.38  & $-$ & 0.70   & 0.58   & 0.09\\
~3~~~\rip\ (mag)                           & 0.48  & 0.07  & 0.36  & $-$ & 0.60   & 0.45   & 0.07\\
~4~~~$M_2$ (\msun)                         & 0.41  & 0.20  & 0.16  & $-$ & 1.00   & 0.32   & 0.14\\
~5~~~$R_2=R_\mathrm{Roche}$ (\rsun)         & 0.73  & 0.12  & 0.54  & $-$ & 1.00   & 0.67   & 0.09\\
~6~~~$f_3$                                 & 2.03  & 0.61  & 1.02  & $-$ & 3.31   & 2.33   & 0.56\\
~7~~~\teffs\ (K)                           & 4119  & 143   & 3918  & $-$ & 4402   & 4185   & 136 \\
~8~~~$-$log($L_2/$\lsun)                   & 0.88  & 0.15  & 1.20  & $-$ & 0.47   & 0.92   & 0.15 \\
~9~~~$M_1$ (\msun)                         & 0.92  & 0.25  & 0.40  & $-$ & 1.30   & 0.88   & 0.26\\
10~~~$q=M_2/M_1$                           & 0.46  & 0.20  & 0.12  & $-$ & 0.85   & 0.38   & 0.18 \\
11~~~$i$ (\degr)                           & 77.3  & 2.9   & 72.9  & $-$ & 85.7   & 78.4   & 2.9 \\
12~~~$a$ ($10^{11}$ cm)                    & 1.64  & 0.16  & 1.24  & $-$ & 1.99   & 1.59   & 0.15\\
13~~~\rci/$a$                              & 0.13  & 0.03  & 0.10  & $-$ & 0.21   & 0.14   & 0.03\\
14~~~\rco/$a$ (\ps$\!=\!3484$\,s)        & 0.20  & 0.01  & 0.19  & $-$ & 0.22   & 0.21   & 0.01\\
15~~~\rco/$a$ (\ps$\!=\!3911$\,s)        & 0.22  & 0.01  & 0.20  & $-$ & 0.24   & 0.22   & 0.01\\
16~~~\rci/\rco\ (\ps$\!=\!3484$\,s)      & 0.65  & 0.09  & 0.54  & $-$ & 0.94   & 0.68   & 0.09\\
17~~~\rci/\rco\ (\ps$\!=\!3911$\,s)      & 0.60  & 0.09  & 0.50  & $-$ & 0.87   & 0.63   & 0.09\\
18~~~$L_\mathrm{acc}$ (\lsun)$^{(1)}$       & 2.10  & 0.79  & 0.89  & $-$ & 5.53   & 1.95   & 0.75\\             
19~~~$\dot M$ ($10^{-10}$\msun/yr)$^{(1)}$  & 6.8   & 4.5   & 1.2   & $-$ & 36.9   & 7.0    & 5.0 \\ 
20~~~$T_\mathrm{wd}$ (kK) $^{(2)}$          & 27.5  & 5.1   & 14.0  & $-$ & 41.3   & 26.3   & 5.0 \\
21~~~\rwd\ ($10^{8}$ cm)$^{(3)}$            & 6.55  & 2.21  & 3.69  & $-$ & 11.86  & 6.93   & 2.23 \\
22~~~$r_\mathrm{wd}$ (mag)$^{(4)}$          & 23.8  & 0.6   & 22.5  & $-$ & 25.2   & 23.7   & 0.6 \\
23~~~$g_\mathrm{wd}$ (mag)$^{(4)}$          & 23.9  & 0.6   & 22.6  & $-$ & 25.4   & 23.8   & 0.6 \\
24~~~$\mu_{34}$ ($10^{34}$ G\,cm)           & 1.61  & 0.29  & 0.90  & $-$ & 2.43   & 1.57   & 0.31\\
25~~~$B_\mathrm{surf}$ (MG)$^{(5)}$         & 100   & 82    & 7     & $-$ & 348    & 85     &  76 \\ 
26~~~$K_1$ (km\,s$^{-1}$)                  &  96   & 32    & 37    & $-$ & 169    & 81     &  27 \\
27~~~$K_2$ (km\,s$^{-1}$)                  & 219   & 34    & 139   & $-$ & 299    & 225     &  37 \\[1.0ex]
 \hline\\          
\end{tabular}\\[-1.0ex]
\footnotesize{ (1)~Based on case-B accretion-induced flux; (2)~Based
  on $\dot M$ in line 19 and Eq.~\ref{eq:temp};
  (3)~Based on $M_1$, $T_\mathrm{wd}$, and WD models with a thick hydrogen envelope
  \citep{althausbenvenuto98,renedoetal10}; (4) Reddened model magnitudes;
  (5) Based on lines 21 and 24.}
\label{tab:sum}    
\end{flushleft}
\end{table}

\subsection{Grid of photometrically accepted models}

We computed a grid of dynamical models of \jshort\ that extended over
$d\!=\!1200-2900$\,pc and \ebmv$=0.37-0.71$ with step widths of 10\,pc
and 0.01, respectively.
Each bin is characterized by the
single value of $M_2$($d$,\ebmv) that matches the observed
$i\!=\!18.98$ (Table~\ref{tab:eclmag}) and contains models for
$M_1$ between 0.40 and 1.30\,\msun\ with a step width of
0.01\,\msun. As side conditions we required $q\!<\!0.85$, which
ensures stable mass transfer, and $M_2\!>\!0.16$\,\msun, which
accounts for the strict lower limit to $M_2$ in evolving CVs with
9\,hr orbital period in the binary population study (BPS) of
\citet{kalomenietal16}, as defined by their Eq. 12. For each
individual model, we calculated a wide range of system
parameters. Results are provided in columns $3\!-\!5$ of
Table~\ref{tab:sum}.
A coarser version of the grid with steps of 0.04 in \ebmv\ and 100\,pc
in $d$ was used for Fig.~\ref{fig:sys}. The left panel shows all
photometrically accepted models in the $d$-\ebmv\ plane as dots or
circled dots. They cover a distance range that stays well within the
Gaia EDR3 confidence limits \citep{bailerjonesetal21}: There are no
accepted models at $d\!<\!1270$\,pc, because $M_2$ would fall below
$0.16$\,\msun, and none at $d\!>\!2500$\,pc, because $q\!>\!0.85$ or
$M_1\,> \!1.3$\,\msun\ would be required. The $M_2$ values quoted in
the figure are the averages of $M_2(d,E_\mathrm{B-V})$ over \ebmv\ for
the selected distance. The quoted spectral types are obtained from
\ebmv\ via \rip\ and do not depend explicitly on $d$.  The center
panel shows the accepted models in the \teffs$-$\logl\ plane, the
Hertzsprung-Russell (HR) diagram of the secondary star. The right
panel shows the theoretical results on the $q$ dependence of the
period ratio \pspo\ together with the observed level in \jshort (green
lines). These results are now discussed in turn.

\subsection{\jshort\ in the Hertzsprung-Russell diagram}
\label{sec:hr}

Using the stellar-locus tables of Mamajek \citep[][see footnote
  5]{pecautmamajek13} and \citet{mannetal15}, we constructed the color
dependence of the effective temperature \teffs\ of the secondary star
as a function of \rip. Each of our models is tagged by a value of
\rip\ and a value of \teffs. Combined with $R_2$ from Eq.~\ref{eq:r2},
we calculated \logl\ for all models and thereby their position in the
HR diagram of the secondary star. The center and left
panels of Fig.~\ref{fig:sys} show the same photometrically accepted
models as small black dots.  The center panel includes, in addition,
results of the BPS of \citet{kalomenietal16}. The red curve indicates
the strict upper limit to the luminosity $L_2$ of the secondary star
as a function of \teffs, taken from their Fig.~20. Our photometrically
accepted models are distributed across this line and for about 50\% of
them the secondary star is too bright by up to a factor of 2.6. The
encircled dots indicate the luminosity-restricted sample of models
that are consistent with the \citet{kalomenietal16} BPS calculations,
allowing for a 10\% uncertainty and spillover in $L_2$. We list the
mean values and standard deviations of all parameters of the
luminosity-restricted sample in columns 6 and 7 of
Table~\ref{tab:sum}. The centroid of this restricted sample is
indicated by the yellow squares and given by
$d\!=\!1596\!\pm\!208$\,pc, \ebmv$\,=\!0.58\!\pm\!0.09$,
$M_2\!=\!0.32\!\pm\!0.14$\,\msun, \teffs$\,=\!4185\!\pm\!136$\,K, and
\logl$\,=\!-0.92\!\pm\!0.15$, where the quoted errors represent the
standard deviations of the distributions and we consider all models a
priori as equally probable. The strongly skewed $M_2$ distribution of
the luminosity-restricted sample still contains $M_2$ values up to
1.0\,\msun, but all models with secondary masses $M_2$ more than
2\,$\sigma$ above the mean, or $M_2\!>\!0.60$\,\msun, lie in the
extended tail of the distribution at $d\!>\!1980$\,pc in the left
panel and at \logl$\,>\!-0.68$ in the center panel of
Fig.~\ref{fig:sys}. This tail contains 5\% of the models. We consider
them correspondingly unlikely and take this as an indication of a
low-mass secondary star in \jshort, with $M_2$ lower than expected for
a normal CV of 9\,hr orbital period.

The two dashed blue curves in the center panel of Fig.~\ref{fig:sys}
show two selected evolutionary tracks of CVs from the BPS calculations
of \citet[][their Figs. 20 and 1]{kalomenietal16}, a normal one with a
secondary near the main sequence that evolves close to the red line
and a binary with an evolved secondary of lower mass that departs from
that track and heads for the lower left of the diagram into the
general region occupied by ultracompact binaries. They represent
possible evolutionary scenarios for \jshort. In the former case,
\jshort\ would probably evolve into a polar, and in the latter, it
might follow the 9~h binary \kic\ of \citet{yuetal19} (magenta dot and
magenta dashed curve) to become an ultracompact
binary, with the decisive difference of one component being highly
magnetic. 

\subsection{Accretion rate and WD temperature} 
\label{sec:aa}

We have no good handle yet on the accretion rate of \jshort, but it is
useful to obtain at least an estimate. Assuming isotropic emission,
the accretion luminosity is $L_\mathrm{acc}\!\simeq\!4\pi
d^2F_\mathrm{acc}$ and the accretion rate is estimated as $\dot
M\!\simeq\!L_\mathrm{acc}R_1/\mathrm{G}M_1$. So far, only the optical
part $F_\mathrm{opt}$ of the accretion-induced high-state flux
$F_\mathrm{acc}$, is available. We identify it with the upper
envelope to the observed (reddened) fluxes in the left panel of
Fig.~\ref{fig:sed}. Integrated from the Balmer edge into the infrared
they give $F_\mathrm{opt,red}\!=\!\ten{2.5}{-12}$\,\ergs. For each
model, we obtained the individually de-reddened value of
$F_\mathrm{opt}$ and estimated
$F_\mathrm{acc}\!=\!(1+f_\mathrm{uvx})\,F_\mathrm{opt}$ with a
correction factor $f_\mathrm{uvx}$ that accounts for the still missing
UV and X-ray fluxes.  We considered two cases, of which case A with
$f_\mathrm{uvx}\!=\!0$ or $F_\mathrm{acc}\!\equiv\!F_\mathrm{opt}$
represents the absolute minimum. Guided by the well-studied IPs
\exhya\ and \vsgr\ \citep{eisenbartetal02,beuermannetal04}, we adopted
case B with $f_\mathrm{uvx}\!=\!2.0$ or $F_\mathrm{acc}\!=\!3.0\times
F_\mathrm{opt}$, accounting for an ultraviolet flux as in other IPs
and a moderate X-ray flux. In what follows we use case B in the
attempt to obtain an overview of the system parameters. The range of
case-B accretion rates is listed in line 19 of Table~\ref{tab:sum}.

The secular mean accretion rate $\langle\dot M\rangle_\mathrm{10}$ of
systems with a sufficiently old WD averaged over its Kelvin-Helmholtz
timescale determines its effective temperature by compressional heating,
\begin{equation}
T_\mathrm{eq}\!=\!18.9\,\langle\dot M\rangle_\mathrm{10}^{1/4}\,M_1~~\mathrm{kK}
\label{eq:temp}
\end{equation}
\citep{townsleygaensicke09}, where $\langle\dot M\rangle_\mathrm{10}$
is in units of $10^{-10}$\,\msunyr\ $M_1$ in solar units. We
estimated the current WD temperature by replacing $\langle\dot
M\rangle_\mathrm{10}$ in Eq.~\ref{eq:temp} with the case-B accretion
rate for each model. The resulting mean temperature of all
photometrically accepted models is 27\,kK (line 15 of
Table~\ref{tab:sum}). The corresponding reddened WD AB magnitudes are
$r\!=\!22.56-25.16$ and $g\!=\!22.57-25.39$, which are all at least
1~mag fainter than our observed 2$\sigma$ upper limits of 21.44 and
21.69, respectively (Sect.~\ref{sec:wd}). The faintest WDs are found in
the most distant models with a primary mass of 1.3\,\msun.

\subsection{Orbital inclination}
\label{sec:incl}

We estimated the inclination of \jshort, using the Roche lobe
parameters $x_1\!=b/a$, $x_4$, $y_4$, $z_6$, and $r_2^*$ in the
notation of \citet{kopal59}\footnote{With the WD at the origin and the
  x axis connecting the two stars, $x_1$ measures the distance to the
  $L_1$ point, $x_4\!\simeq\!1$ locates the position along the
  $x$ axis of maximum excursion of the lobe, $y_4$ is the lobe
  position in the orbital plane and $z_6$ in the direction of the
  rotational axis, and $r_2^*$ is the equivalent volume-filling radius
  of the secondary star.}, expressed as functions of the mass ratio
$q$. For a given duration of the eclipse in phase units, the orbital
inclination is a function of $q$ \citep[e.g.,][]{horne85}. For
\jshort, the inclination is 87\fdg6, 81\fdg8, 77\fdg4, or 73\fdg3 for
$q\!=\!0.1, 0.2, 0.4$, or 0.8, respectively.  The path of the WD
across the shadow of the secondary star defines an arced section with
a chord length $\rho$ and a height $\sigma$, which are also functions
of $q$. For the luminosity-selected sample in Fig.~\ref{fig:sys}
(encircled dots), the mean inclination is 78\fdg4 and the full range
is $72\fdg9\!-\!85\fdg7$ (Table~\ref{tab:sum}). At eclipse center, the
WD has dived into the shadow of the secondary star to a mean depth
$\sigma\,=\!20.6$\rwd\ (full range $6.2\!-\!58.2$\rwd). The
magnetically guided part of the accretion stream reaches a similar
height above the orbital plane and, depending on the details of the
accretion geometry, only a small part of the stream emission escapes
eclipse. Hence, the system qualifies as deeply eclipsing.

\subsection{The magnetic moment of the WD}
\label{sec:mag}

In an IP, matter accretes to the magnetic field of the WD near the
magnetospheric radius, \rmag, which is usually quoted as a fraction of
the Alfv{\'e}n radius, \ralf. The latter is obtained by equating the
ram pressure of the infalling matter for isotropic accretion with the
magnetic pressure of a dipolar field.  The rapid drop of the magnetic
pressure with increasing radius as $r^{-6}$ implies that the usual
estimate of \rmag$\,\simeq\!0.5\,$\ralf\ cannot be severely
wrong. The factor 0.5 accounts for the difference between spherical
and equatorial accretion \citep{franketal02}, and results in
\begin{equation}
  r_\mathrm{mag}\,\simeq\,\ten{2.2}{8}\dot
  M_{10}^{-2/7}M_\mathrm{wd}^{-1/7}\mu_{30}^{4/7} ~~~\mathrm{cm}.
 \label{eq:rmag}
\end{equation}
For each of our models of \jshort, we obtained an estimate of the
magnetic moment $\mu_{30}$ of the WD in units of $10^{30}$\,G\,cm$^3$
by using the case-B accretion rate in units of $10^{-10}$\,\msunyr,
the WD mass in solar units, and equating \rmag\ with the corotation
radius \rco$\,=\!(GM_1/\omega_\mathrm{s}^2)^{1/3}$.  For the sample of
models restricted by the photometry and luminosity of the secondary
star (encircled dots in Fig.~\ref{fig:sys}), we found a mean magnetic
moment $\mu\!=\!\ten{(1.6\,\pm\,0.3)}{34}$\,G\,cm$^3$
(Table~\ref{tab:sum}, line 24), which is at the upper limit of the
range discussed for IPs \citep{nortonetal04} and is
more characteristic of a polar. The implied field strength would be
typical of or exceed that of a polar as well.

\subsection{Spin equilibrium}
\label{sec:eq}

Most IPs are expected to accrete in spin equilibrium \mbox{($\dot
  P_\mathrm{spin} = 0$)}, because a mismatch between the velocities of
matter and field at the accretion radius \racc\ leads to rapid spin-up
or spin-down of the WD, ensuring
\rmag$\,\simeq\,$\racc$\,\simeq\,$\rco\ most of the time.  Matter
leaving $L_1$ with a specific angular momentum
$j_\mathrm{L1}\!=\!b^2\Omega_\mathrm{o}$ will settle into a Keplerian
orbit at the circularization radius \rci\ with
$j_\mathrm{ci}\!=\!(\mathrm{G}M_1r_\mathrm{ci})^{1/2}\!=\!j_\mathrm{L1}\zeta$,
giving \rci$\,=\!(1\!+\!q)(b/a)^4\zeta^2$. Here, $b$ is the separation
between the WD and $L_1$, $\Omega_\mathrm{o}\!=\![GM_1(1\!+\!q)/a^3]^{1/2}$
is the orbital frequency, and $\zeta(q)\!\simeq\!0.84\!-\!0.89$ for
$q\!=\!1.0\!-\!0.1$, respectively, is the fraction of the angular
momentum at $L_1$ that is preserved at \rci, when the pull of the
secondary star or the non-sphericity of the problem is accounted for
\citep{flannery75,lubowshu75}.
For \rmag$\,\simeq\,$\rco$\,\ga\,$\rci, a viscous disk cannot form
and the WD accretes from the stream, either directly or after
circularization. This is obviously the case in \jshort: the
entire sample of our dynamical models that match the photometry of
\jshort\ (dots and circled dots in Fig.~\ref{fig:sys}) have
\rco$\,>\,$\rci\ (Table~\ref{tab:sum}, lines 13 to 17).

\citet{kinglasota91} argued that the accretion of
$j_\mathrm{ci}\!=\!j_\mathrm{L1}\zeta$ in spin equilibrium implies
\rco$\,\simeq\,$\rci\ as well, leading to
\begin{equation}
P_\mathrm{spin}/P_\mathrm{orb} = \Omega_\mathrm{o}/\omega_\mathrm{s} \simeq (1+q)^2(b/a)^6\zeta^3,
\label{eq:kl91}
\end{equation}
where $b/a\!\simeq\!0.500\!-\!0.227$log\,$q$
\mbox{\citep{plaveckratochvil64}}. They identified the preferred
occurrence of IPs near \pspo$\,=\!0.10$ with stream-fed accretion in
spin equilibrium. A period ratio of 0.10, however, or larger as in
\jshort, requires $q\!\la\!0.15$, which is much lower than the typical
$q\!\approx\!0.6$ for nine IPs above the period gap that have measured
component masses \citep{ritterkolb03}. The likely cause is that
\rco\ and \rci\ do not agree as closely as assumed.  As a consequence,
a reliable value of $q$ for \jshort\ cannot be obtained from
Eq.~\ref{eq:kl91}.

The relation between \rco\ and \rci\ in spin equilibrium is determined
by the exchange of angular momentum between matter and field inside
and outside \rco, which may take place over a substantial radial
interval. An example is the accretion from a viscous disk. Another is
the accretion of a stream of diamagnetic blobs of matter studied
analytically by \citet{king93} (henceforth K93) and numerically by
\citet{wynnking95} (henceforth WK95). Such blobs plunge deep into the
magnetosphere on quasi-ballistic orbits, experience a drag by the
excitation of Alfv\'en waves, when they move across field lines and
``pluck them like violin strings'' \citep{drelletal65}, until they are
broken up by Kelvin-Helmholtz instabilities, invaded by the field, and
accreted \citep{aronslea80}. K93 made the problem analytically
tractable by treating the blobs as test particles, disregarding
changes in their internal state, neglecting viscous collisions between
them, assuming spherical symmetry ($\zeta\!=\!1$), selecting a simple
parameterization of the drag, and describing the blob orbits as Kepler
ellipses that slowly change under the influence of the drag. He
described the drag as $\vec{F}\!=\!-k_0\,r^{-n}\vec{\upsilon}$, where,
$\vec{\upsilon}$ is the velocity of the blob relative to the field and
the integer $n\!\ge\!0$ regulates the radial dependence of the
drag. The azimuthal motion of a blob is typically accelerated near
apastron and decelerated near periastron, which circularizes the orbit.
K93 expressed the equilibrium period ultimately as a function of $n$
and $q$ (his Eqs. 38 ff). The relation between the characteristic
radii becomes \rci$\,=\!(f/g)^{2/3}\,$\rco, with functions $f(q)$ and
$g(q)$ for a given parameter $n$. This relation bears a formal
similarity to $r_\mathrm{in}\!=\!\omega_\mathrm{s}^{2/3}r_\mathrm{co}$
for disked systems \citep{wang95}, with $\omega_\mathrm{s}$ the
fastness parameter (not to be mixed up with the spin angular
velocity). In both cases a competition exists between orbits that
accelerate the magnetosphere inside \rco\ and decelerate it outside
\rco. The decisive dissimilarity is the absence of viscous inter-blob
interactions in the model of K93 and WK95, which holds as long as the
blobs do not accumulate and form a nascent disk. Consequently, the
radial transport of angular momentum, characteristic of an accretion
disk, is absent in the model of K93 and leads to a relation
between \pspo\ and $q$ (his Eq.~37), 
\begin{equation}
  P_\mathrm{spin}/P_\mathrm{orb} = (1+q)^2~(b/a)^6\,(g/f)~\zeta^3,
\label{eq:k93}
\end{equation}
where we have added the factor $\zeta^3$ to correct approximately for
the neglect of the gravitational pull of the secondary star. The
functions $f(q)$ and $g(q)$ are given by K93 and WK95 for $n\!=\!0, 2$,
and 3. For $n\!=\!1$, $g/f\!=\!1$. The numerical study of WK95,
treating the particle orbits correctly in the nonspherical geometry,
gave period ratios that consistently exceeded the analytical results,
less so for $n\!=\!0$ than for positive~$n$. Unfortunately, only two
sets of results for \pspo\ versus $q$ were published.

In the right panel of Fig.~\ref{fig:sys}, we compare the
$q$ dependence of \pspo\ obtained from the analytical and numerical
calculations of K93 and WK95 for $n\!=\!0$ to 2 with the level of the
observed \pspo\ of \jshort\ (green lines). As noted in
Eq.~\ref{eq:k93}, the correction $\zeta(q)$ from \citet{lubowshu75}
has been applied to the analytical period ratios, while the numerical
ones correctly account for the non-sphericity of the gravitational
field. We note that the numerical values of $\zeta$ displayed in
Fig.~1 of \citet{kingetal90} show the opposite trend from those of
\citet{lubowshu75}, which leads to a period ratio in WK95 that varies
little with $q$. The numerical results exceed the analytical
ones by about 30\% for $n\!=\!0$ and by more for $n\!=\!2$. The
analytical $n\!=\!0$ model and the numerical $n\!=\!0$ and $n\!=\!2$
models best match the observed period ratio of \jshort. Hence, the K93
and WK95 models support the nature of \jshort\ as a stream-fed IP in
spin equilibrium, but they are not sufficiently accurate to allow a
determination of $q$ from \pspo.

\section{Discussion}
\label{sec:dis}

\jmid\ is the first deeply eclipsing stream-fed IP. It is disk-less in
the sense that it lacks a viscous (i.e., luminous) accretion disk. It
combines two properties that are extremely rare within the
class. There are only five IPs that display deep eclipses of the
central accretion region surrounding the WD, the two long-known
systems DQ Her and XY Ari, V902~Mon=IPHAS\,J0627 \citep{aungwer12},
Nova\,Sco~1437 \citep[][and references therein]{potterbuckley18}, and
V597~Pup, the now faint remnant of Nova Puppis 2007
\citet{warnerwoudt09}, and all of them accrete via disks. Stream-fed
systems are even less common and none of the candidates is eclipsing.
The identification of a stream-fed system may be ambiguous because of
the phenomenon of stream overflow that occurs also in disked systems
\citep{lubow89}.  The best and so far only example of a stream-fed IP is
\vv\ \citep{buckleyetal95,buckleyetal97,hellierbeardmore02,joshietal19},
which, however, is seen at an inclination of only $i\!\sim\!10$\degr,
preventing a measurement of its component masses. The combination of a
deep eclipse and a stream-fed and seemingly disk-less accretion
geometry in \jshort\ is unique with important prospects for follow-up
studies.

Our Fourier and periodogram analyses gave a clear signal at a single
period \pobs$\,=\!3911$\,s. Our data contain no hint for identifying
it with either the spin or the synodic period. If a stream of tenuous
matter impinges on the magnetosphere, is instantaneously permeated by
the field, and accreted on the spot, the resulting signal will be
pulsed at the synodic (beat) period. Such matter is generally hold
responsible for the emission of hard X-rays. Denser blobs of matter,
on the other hand, continue on ballistic orbits in the magnetosphere
and are carried around by the field until threaded. They lose the
memory of the orbital phase of impact and the resulting emission is
likely pulsed on the spin period. For a structured stream, covering a
wide range of densities, a mixture of spin and synodic periods and
their side bands may be expected \citep{warner86}, and a prediction
for an individual wavelength band is not easily made.
In the stream-fed IP \vv, the X-ray flux is pulsed at
the synodic and the spin period \citep{buckleyetal97,joshietal19}, the
optical circularly polarized flux appears at the spin period
\citep{buckleyetal95}, and the optical continuum preferentially at the
synodic period \citep{hellierbeardmore02}. Performing a hard X-ray
study and optical circular spectropolarimetry are, therefore, the most
obvious next steps for further study of \jshort.

There is preliminary, and certainly incomplete, evidence that
\jshort\ may not be a strong X-ray source (Sect.~\ref{sec:sec2}). With
an interstellar atomic hydrogen column density
\nh$\,\simeq\!\ten{5}{21}$\,\atoms, based on \ebmv$\,\simeq\!0.54$
\citep{nguyenetal18}, X-rays with $E\!<\!1$\,keV would be absorbed. A
strong soft X-ray component was observed in some IPs (e.g., PQ~Gem,
V405~Aur, and UU~Col), and such a source would not be detectable in
\jshort. The spectroscopic detection of \heii\ line emission would,
however, demonstrate the presence of such a source. Other IPs emit
hard X-rays, possibly reflecting differences in the accretion geometry
and in the density structure of the accreted matter. For example,
Nova\,Sco~1437 and XY~Ari are strong hard X-ray sources, while
IPHAS\,J0627 is faint. \citet{aungwer12} argued that the X-ray
emission region in IPHAS\,J0627 may be obscured by the inner edge of
the disk, but this may not be the whole story.

The lack of a luminous disk and the observed stream flipping in
\jshort\ are not short-lived transitory phenomena. The evidence for
the stream flipping is based on the pulse phase dependence of the
eclipse centers and the orbital phase dependence of the pulse minima
observed in all three observing seasons (Sect.~\ref{sec:flip}).  These
results strongly suggest that \jshort\ is stream-fed, a conclusion
that is supported also by the fact that all of our photometrically
accepted models have \rco$\,>\,$\rci\ and, hence, lack a classical
viscous disk.  The accretion process is regulated by the hierarchy
between three characteristic timescales, the dynamical (Kepler)
timescale \tkep, the threading timescale \tmag, and the viscous
timescale \tvis. On-the-spot accretion takes place for
\tmag$\,\ll\,$\tkep, while the matter is circularized before being
accreted for \tmag$\,\gg\,$\tkep, assuming in both cases that
\tvis\ is much longer. A theory for the former case that allows for a
structured stream is not available. A theory for the latter, assuming
the accretion of dense diamagnetic blobs of matter, was devised by K93
and WK95 and discussed in Sect.~\ref{sec:eq}. In their model, the
matter is circularized by the drag it experiences when crossing field
lines. It does not spread into a disk as long as \tmag\ stays
sufficiently short and the blobs do not accumulate and get viscously
into contact.  Because of the similarities with disk theory, K93
insisted on calling his accretion mode stream-fed and not disk-less. A
relation of the form of Eq.~\ref{eq:k93} exists for stream-fed, but
not for disk-fed systems. It is, however, burdened by the factor $g/f$
in the theory of K93 or a \mbox{corresponding factor in any
  alternative theory}.

The factual absence of viscous interaction would render any ring-like
structure cool. On the other hand, if it is irradiated and heated by a
central X-ray source, it may be difficult to hide it at optical
wavelengths. In any case, the presence of Balmer and helium line
emission may help to trace the geometry of the magnetically guided
stream and the potential presence of an accretion ring.
K93 estimated that individual blobs may survive in the vicinity of
\rco\ for up to ten orbits around the WD, before they lose their
identity, are invaded by the field, and accreted.  We estimated the
times scales $t_\mathrm{drag}$ of \citet{drelletal65} and K93 and
$t_\mathrm{evap}$ of \citet{aronslea80} for the parameters of
\jshort\ and obtained smaller numbers comparable to one pulse
period of the WD, which is not enough to establish a ring, but
sufficient for the matter to lose the memory of the impact point.
Whether a well developed ring exists in \jshort\ and whether the
synodic or spin period is preferred for the optical continuum emission
are, therefore, open questions.

Of the stellar components, we have a secure photometric detection of
the secondary star at the center of the eclipse, with a Pan-STARRS
i-band AB magnitude of 18.98(3) and a spectral type $\sim\,$K6. The
mean $M_2$ of the models that fit the photometry (Table~\ref{tab:sum},
column 7) is $0.32$\,\msun\ and the 2$\sigma$ range is
$0.16\!-\!0.60$\,\msun. The secondary star is bloated over a main
sequence star by a factor $f_3\!=\!1.4\!-\!3.3$
(Table~\ref{tab:sum}). This finding suggests that \jshort\ may have
previously passed through a phase of thermal timescale mass transfer,
which would determine its future evolution as well. From its present
position in the HR diagram of the secondary star (Fig.~\ref{fig:sys},
center panel), the system may either still evolve into a polar or end
up as an ultracompact binary, as predicted also for the nonmagnetic
CV \kic\ \citep{yuetal19}, but in the present case one of the final
degenerate components would be strongly magnetic.

Detecting the WD in \jshort\ will not be a trivial task, given the
2$\sigma$ upper limits to its reddened (observed) magnitude of
$g\!>\!21.69$ and $r\!>\!21.44$ (Sect.~\ref{sec:wd}). Our dynamical
models suggest probable magnitude ranges of $g\!=\!23.8\!\pm\!0.6$ and
$r\!=\!23.7\!\pm\!0.6$, but the WD may reach 25 mag in both colors if
the WD mass is as high as 1.3 \,\msun. The quoted magnitudes are based
on the radii of WDs with a thick hydrogen envelope
\citep{althausbenvenuto98,renedoetal10} and effective temperatures
computed from Eq.~\ref{eq:temp} with the long-term accretion rate
replaced by the current case-B rate (Sect.~\ref{sec:aa}). They
disregard accretion-heated hot polar caps and can only be
approximate. The dK6 secondary star with $r\!=\!18.98$ can be traced
spectroscopically by its photospheric absorption lines or by emission
lines from its face heated by the WD. Given the frequent low states, a
phase-resolved study appears feasible. Tracing the motion of the WD
will be more challenging. The predicted velocity amplitudes $K_1$ and
$K_2$ of primary and secondary star are quoted in Table~\ref{tab:sum}.

The estimated magnetic moment of the WD in \jshort\ of
$\mu\!\simeq\!\ten{1.6}{34}$\,G\,cm$^3$ suggests a surface field
strength typical of polars \citep{nortonetal04}. Hence, studying the
magnetic field structure by means of spectropolarimetry and cyclotron
spectroscopy seems a realistic prospect. Besides the high magnetic
moment, the comparatively low accretion rate in the high state,
assuming case-B to be valid, is reminiscent of long-period polars
\citep{townsleygaensicke09}. While low in comparison with nonmagnetic
CVs, the case-B accretion rate substantially exceeds that expected
from gravitational radiation.

\section{Outlook}
\label{sec:summary}

We have presented a detailed photometric study of the first deeply
eclipsing stream-fed IP that demonstrably lacks a viscous accretion
disk. The $grizy$ photometry through the total eclipse yielded the
magnitude and spectral type of the moderately evolved secondary star,
while the WD has so far remained undetected. The Milky Way object is
tolerably absorbed at a distance of about 1.6\,kpc and is observable from
both hemispheres. Key follow-up observations include: (i) a hard X-ray
study and optical circular polarimetry to identify the spin period;
(ii) phase-resolved optical spectroscopy to measure the velocity
structure of the stream and the HeII$\lambda$4686 emission line
flux; (iii) high-cadence high-state blue photometry through the
eclipse to obtain information on the accretion geometry; (iv) the same
in a deep low state to measure the diameter and mass of the WD; (v)
high-resolution spectroscopy of the K star in a low state; and (vi)
circular spectropolarimetry to measure the strength and structure of the
magnetic field of the WD. A dedicated theoretical study is needed to
understand its past and future evolution.


\begin{acknowledgements}
  \noindent
  We thank the anonymous referee for a very helpful report that
  improved the paper. We thank Axel Schwope for reading an earlier
  version of the paper and providing useful comments, Tim-Oliver
  Husser for essential help with using the MONET/S telescope for this
  project, and Boris Gaensicke for help in accessing the UKIRT
  archive. We thank the European Space Agency (ESA) for supporting our
  project within the Near-Earth Segment of the Space Situational
  Awareness Program of ESA under contract number 4000116155/15/D/AH
  (P2-NEO-VIII), by assigning observation time at the Calar Alto
  Schmidt telescope. We thank the Faulkes Telescope Project,
  Cardiff/UK, as partner of the Las Cumbres Observatory (LCO) for
  providing telescope time without much bureaucracy on sometimes very
  short notice. We also thank the LCO for Director's Discretion Time
  for using MuSCAT3 at the Faulkes Telescope North on Maui, HI, an
  instrument that was developed by the Astrobiology Center under
  financial supports by JSPS KAKENHI (JP18H05439) and JST PRESTO
  (JPMJPR1775). Part of the photometric data were collected with the
  MONET/S telescope funded by the Alfried Krupp von Bohlen und Halbach
  Foundation, Essen, and operated by the Georg-August-Universit\"at
  G\"ottingen and the South African Astronomical Observatory.  We made
  use of the PanSTARRS data base, the UKIRT Galactic Plane Survey, and
  further sources accessed via the VizieR Photometric viewer operated
  at CDS, Strasbourg, France.

\end{acknowledgements}

\bibliographystyle{aa}

\end{document}